\documentclass[fleqn,usenatbib]{mnras}

\usepackage{newtxtext,newtxmath}
\usepackage[T1]{fontenc}

\DeclareRobustCommand{\VAN}[3]{#2}
\let\VANthebibliography\thebibliography
\def\thebibliography{\DeclareRobustCommand{\VAN}[3]{##3}\VANthebibliography}

\usepackage{graphicx}	% Including figure files
\usepackage{amsmath}	% Advanced maths commands
\usepackage{pdflscape}
\usepackage{booktabs}
\usepackage{multirow}
\usepackage{orcidlink}

\defcitealias{2003sirko}{SG03}
\defcitealias{TQM05}{TQM05}

\title[Self-Regulated Accretion in AGN Disks]{Jet Feedback and the Self-Regulated Growth of Black Holes \\ Embedded in AGN Disks}

\author[Epstein-Martin, Su, Haiman, \& Perna]{
Marguerite Epstein-Martin\orcidlink{0000-0001-9310-7808},$^{1}$
Kung-Yi Su\orcidlink{0000-0003-3729-1684}$^{2,3}$
Zolt\'an Haiman\orcidlink{0000-0003-3633-5403}$^{4,1,5}$\thanks{E-mail: zoltan.haiman@ista.ac.at }
and Rosalba Perna\orcidlink{0000-0002-3635-5677}$^{6}$
\\
$^{1}$Department of Astronomy, Columbia University, 550 West 120th Street, New York, NY 10027, USA\\
$^{2}$Department of Physics \& Astronomy and CIERA, Northwestern University, 1800 Sherman Ave, Evanston, IL 60201, USA \\
$^{3}$Black Hole Initiative, Harvard University, 20 Garden Street, Cambridge, MA 02138, USA \\
$^{4}$Institute of Science and Technology Austria (ISTA), Am Campus 1, Klosterneuburg 3400, Austria\\
$^{5}$Department of Physics, Columbia University, 550 West 120th Street, New York, NY 10027, USA\\
$^{6}$Department of Physics and Astronomy, Stony Brook University, Stony Brook, NY 11794-3800, US
}

\pubyear{\the\year{}}

\date{Accepted XXX. Received YYY; in original form ZZZ}

\pubyear{\the\year{}}

\begin{document}
\label{firstpage}
\pagerange{\pageref{firstpage}--\pageref{lastpage}}
\maketitle

%%%%%%%%%%%%%%%%%%%%%%%%%%%%%%%%%%%%%%%%%%%%%%%%%%%%%%%%%%%%%%%%%%%%%%%%%%%%%%%%%%%%%%
%%%%%%%%%%%%% ABSTRACT %%%%%%%%%%%%%
%%%%%%%%%%%%%%%%%%%%%%%%%%%%%%%%%%%%%%%%%%%%%%%%%%%%%%%%%%%%%%%%%%%%%%%%%%%%%%%%%%%%%%

\begin{abstract}

Accretion disks in active galactic nuclei (AGN) are expected to host embedded stellar-mass black holes (BHs) whose gas capture rates can be highly super-Eddington. Without feedback, they would rapidly grow and deplete the disk, but at such high feeding rates, radiation is advected inward with the flow.  We test whether the resulting accretion-powered jet and its shocked cocoon can (i) self-regulate the BH's growth rate and (ii) break out of the AGN disk.
Using \textsc{gizmo}, we perform three-dimensional simulations of a $10\,{\rm M}_\odot$ BH, embedded in dense AGN-disk gas, launching a particle-spawned jet whose mass flux scales with the accretion rate. We explore a range of jet velocities, ambient temperatures and densities, and three cooling regimes.
We find that the jet-inflated cocoon stalls near the Bondi radius and maintains a modest aspect ratio and the self-regulated mass flux is determined by momentum balance at the Bondi radius. Cooling controls the mode of self-regulated accretion, from large-amplitude oscillation cycles at low metallicity to weak, quasi-steady modulation at solar metallicity.
Scaling our results to outer disk conditions implies that mechanical feedback limits BH growth to within an order of magnitude of the Eddington rate -- largely preventing the overgrowth problem.  Over the same disk region, the cocoon remains confined within the disk.  
In the inner disk, accretion onto the embedded BHs
is limited by tidal shear rather than the Bondi radius, and our simulations no longer apply. Whether runaway growth and cocoon breakout remains possible in these inner regions needs further study.

\end{abstract}


%%%%%%%%%%%%%%%%%%%%%%%%%%%%%%%%%%%%%%%%%%%%%%%%%%

%%%%%%%%%%%%%%%%% BODY OF PAPER %%%%%%%%%%%%%%%%%%

% ----------------------------------------------------------------------------------------------------
% Introduction
% ----------------------------------------------------------------------------------------------------

\section{Introduction}\label{sec:introduction}

Accretion disks in active galactic nuclei (AGN) are increasingly recognized as dynamic environments capable of hosting a substantial population of embedded stars and compact objects (COs), including stellar-mass black holes (BHs). Such embedded populations may arise through dynamical capture of stars and stellar remnants from the surrounding nuclear star cluster \citep{1983ostriker,1991syer}, as well as through in situ star formation in the self-gravitating outer regions of the AGN disk \citep{2003goodman,2007nayakshin}. As massive embedded stars evolve off the main sequence, they are expected to leave behind neutron stars and BHs, further enhancing the compact object population within the disk \citep{2003levin-a, 2007levin, gilbaum2022, epstein-martin2025, 2017stone}.

Growing observational and theoretical evidence supports the existence of these embedded populations. In particular, AGN disks have emerged as promising environments for explaining several otherwise unexpected properties of some of the gravitational wave (GW) events detected by the LIGO-Virgo-Kagra (LVK) collaboration. The unusually high total mass of GW190521---whose primary component falls within the pair-instability mass gap \citep{2020abbott190521}--- has been interpreted within the context of hierarchical assembly and gas-assisted mergers in AGN disks \citep{2019mckernan,2020tagawa, 2026bartos}, as has the more recent GW231123 \citep{2025delfavero,2026bartos}, the most massive binary BH merger reported to date \citep{2025abac}. More broadly, the dense, dissipative environment of AGN disks naturally facilitates the formation, hardening, and merger of compact-object binaries \citep{2012mckernan, 2014mckernan, 2018mckernan, 2017bartos, 2017stone, 2020tagawa, 2024vaccaro}, potentially accounting for features such as high component masses \citep{2019yang, 2026vaccaro} and residual eccentricities \citep{2022samsing,2021tagawa} reported in recent GW observations.

A merger embedded in dense gas may be accompanied by an electromagnetic (EM) counterpart, supplying a distinct observational signature and an additional window into the AGN embedded BH population \citep{2017bartos, 2017stone, 2026ford}. Several candidates have been reported including an optical flare following GW190521 \citet{2020graham}, and further optical candidates since \citet{2023graham}, gamma-ray transients coincident with GW triggers \citep{2016bagoly, 2016connaughton}, and a hard X-ray flare following S241125n \citep{2024delaunay,2024wang}, although the specific associations remain contested \citep{2021ashton,2021palmese,2025veronesi,2025he}, since the optical identifications rest on light-curve shape and magnitude criteria that do not by themselves isolate an AGN-disk origin. Still, in the models proposed for these flares the emission is powered by a jet launched from a merger remnant and shocking the surrounding disk gas \citep{2023tagawa_highE, 2026tagawa, 2026chen}, so that the outflow-driven shocks invoked to make mergers observable are of the same kind as those we examine here for a solitary embedded BH.

Gas accretion onto embedded COs is an inevitable consequence of their immersion within the AGN disk. For typical disk conditions, the mass capture rate is predicted to be highly super-Eddington \citep{2017stone}. If such rapid accretion were sustained without regulation, embedded COs would grow quickly, significantly perturbing or even depleting the surrounding disk \citep{2004goodman,  2007levin, 2012mckernan, 2022tagawa}. To avoid this outcome, many analytic AGN disk models cap accretion onto embedded objects at the Eddington rate \citep{2020tagawa, gilbaum2022, epstein-martin2025}. While this assumption preserves the global structure of the disk, it is not strongly motivated by the physical processes expected to operate under highly super-Eddington conditions, where radiation is trapped and advected inward rather than escaping to provide feedback \citep{1977katz, 1978begelman, 1988abramowicz}.

In practice, accretion onto embedded BHs is more plausibly regulated by feedback from the accretion flow itself. One such mechanism is radiative feedback: radiation pressure escaping from the geometrically and optically thick super-Eddington circum-BH disk (CBD) can drive gas ejection and, in principle, regulate the accretion flux \citep{2009milosavljevic_a, 2009milosavljevic_b, 2011park, 2012park}. In the AGN context, however, this mechanism is insufficient to solve the parallel problems of BH overgrowth and disk depletion, as the ejected gas is driven preferentially perpendicular to the disk plane \citep{2017sugimura, 2021toyouchi}, leaving the in-plane inflow that feeds the BH largely intact, and even an isotropic radiation field has been shown to suppress accretion only modestly \citep{2016inayoshi}.

Mechanical feedback is a more efficient driver of ejection than radiative feedback, typically understood to function through a duty cycle in which rapid accretion drives mass-loaded outflows that clear gas from around the central object and lower its time-averaged accretion rate. In this scenario, the outflow is decelerated where it meets the surrounding medium, forming a working surface at which a reverse shock halts the outflow material and a forward shock sweeps up ambient gas ahead of it. Shocked outflow material accumulates around the outflow, forming a hot, low-density cocoon \citep{1974scheuer, 1989begelman}. Over-pressured with respect to the ambient medium, the cocoon expands and evacuates the surrounding material, thereby reducing the supply of cool, dense gas available to the central object. In suppressing accretion, the cocoon also reduces or extinguishes the driving outflow. Accretion resumes only after the ambient medium refills the evacuated cavity. This self-regulated accretion process is sometimes termed the jet feedback mechanism (JFM), and has been identified across a broad suite of astrophysical contexts including galaxy clusters \citep{2012mcnamara}, galaxy formation \citep{2012fabian}, planetary nebulae \citep{2002balick}, common-envelope evolution \citep{2017moreno, 2019lopez, 2021grichener}, and young stellar objects \citep[see][]{2016soker}.

The AGN disk environment presents several complicating additions to the classic JFM. In particular, the limited vertical extent of the disk bifurcates the outflow behavior depending on whether or not the cocoon breaks out of the disk. Once the cocoon reaches the disk surface its interior depressurizes and the shocked outer shell accelerates down the vertical density gradient, changing both the rate of subsequent expansion and the timescale on which the evacuated cavity refills \citep{2021kimura, 2023chen}. Breakout therefore sets the duty cycle that determines the time-averaged growth of the embedded BH, and may also convert an otherwise hidden feedback cycle into an observable one. Radiation trapped behind the shock escapes once the optical depth ahead of it falls to $c/v_{\rm sh}$, producing thermal X-ray flares \citep{2023tagawa_solitary, 2025chen}. Overlying gas reprocesses whatever escapes, with broad-line clouds scattering soft X-rays over days to weeks and the dusty torus re-radiating in the infrared over months to years \citep{2021kimura}.

A second complication arises from the differential rotation of the disk, which becomes dominant where the tidal field of the SMBH rather than the gas sound speed set the geometric bounds of the accretion region. Inflow then arrives carrying the specific angular momentum imparted by the shear and circularizes into a rotationally supported flow rather than falling radially. The same shear also acts on the outflow. Once the expanding cavity slows to the local shear velocity, the background flow deforms it azimuthally, stretching the cocoon along the direction of gas flow \citep{1995rozyczka, 2021moranchel}.

Two analytic studies have addressed jet/cocoon feedback in the AGN environment directly. \citet{2022tagawa} considered feedback driven both by relativistic jets from rapidly spinning BHs \citep{1977blandford, 2023kaaz} and by sub-relativistic anisotropic winds \citep{2014jiang, 2015sadowski}, working in the adiabatic limit. They find that in the relativistic case, the jet/cocoon vents before the accretion cycle completes. The evacuated region therefore never grows much larger than the disk is thick, and the disk thickness in turn fixes how long refilling takes. Because refilling is far slower than the draining of the bound CBD, the BH is quiescent for most of the cycle and the time-averaged accretion rate is reduced by a factor of $\sim 10$--$100$. The slower wind shock instead decelerates before reaching the disk surface, so that it remains confined long enough to cool, and gives comparable suppression only beyond a distance from the SMBH of $a \sim 0.1$~pc where radiative losses are inefficient. Shear does not enter their cocoon dynamics, since the evacuated region remains smaller than both the disk scale height and the Hill radius.

A parallel study by \citet{2023chen} treats the same problem for an isotropic wind,  arguing that the wind isotropizes before it can sweep up appreciable disk gas, so that the structure it inflates is a spherical cavity rather than an elongated cocoon. As in \citet{2022tagawa}, breakout arrests the vertical expansion and the cavity thereafter spreads laterally along the disk, and they likewise find that the shell reaches the surface before accretion ceases across most of the conditions they sample. Suppression factors of $\sim 0.005$--$0.06$ and $\sim 0.02$--$0.3$ follow in the adiabatic and radiative limits respectively. Comparing the cavity half-width their unsheared model predicts against the size at which the shear velocity exceeds the cocoon expansion velocity, and finding the two comparable, they argue that shear should markedly deform the cavity without substantially altering the accretion suppression factor.

Numerical work on a related system suggests a different regulating mechanism. \citet{2023su, 2025su} simulate accretion-powered jets and winds from massive BH seeds in the dense, low-metallicity gas of high-redshift protogalaxies, with the outflow scaling self-consistently with the accretion rate rather than being externally imposed. They find the time-averaged accretion rate set by momentum balance between the inflow and the outflow, so that what limits growth is the feedback's ability to overcome the ambient ram pressure rather than a duty cycle of evacuation and refilling.

The propagation of a jet- or wind-driven cocoon in an AGN disk has not, to our knowledge, been simulated directly. In this work we do so, targeting the outer-disk conditions where accretion suppression is predicted to be substantial for sub-relativistic outflows \citep{2022tagawa} and where in-situ star and compact-object formation should occur. We find that jet feedback regulates accretion at the Bondi scale, confirming in the AGN-disk regime the momentum-balance picture of \citet{2023su, 2025su}, and holding the BH growth rate within roughly an order of magnitude of the Eddington accretion rate. Moreover, we find that cocoons do not expand self-similarly but evolve close to the Bondi radius, almost always preventing breakout from the disk.

The rest of this paper proceeds as follows: \S\ref{sec:AGN_params} sets out the AGN conditions and parameter regime of interest, together with the accretion rates and outflow strengths they imply. \S\ref{sec:theory} summarizes the analytic framework for isotropic \citep{1975castor, 1977weaver, 1992koo_a, 1992koo_b} and elongated \citep{1989begelman, 2011bromberg} cocoon propagation, and its extension to self-regulating feedback \citep{2021su, 2023su, 2025su}. \S\ref{sec:methods} describes the numerical methods. In \S\ref{sec:simulation_results} we present the simulations, compare them against the analytic framework, and fit updated feedback scaling relations for embedded BHs. \S\ref{sec:discussion} applies those relations across the broader AGN parameter space, estimating BH growth rates and cocoon breakout. Finally \S\ref{sec:conclusions} summarizes our main conclusions and their implications. 

% ----------------------------------------------------------------------------------------------------
% Sec: AGN_Params
% ----------------------------------------------------------------------------------------------------

\section{AGN parameter space and regimes of interest}\label{sec:AGN_params}

\begin{figure*}
    \centering
    \includegraphics[width=0.98\linewidth]{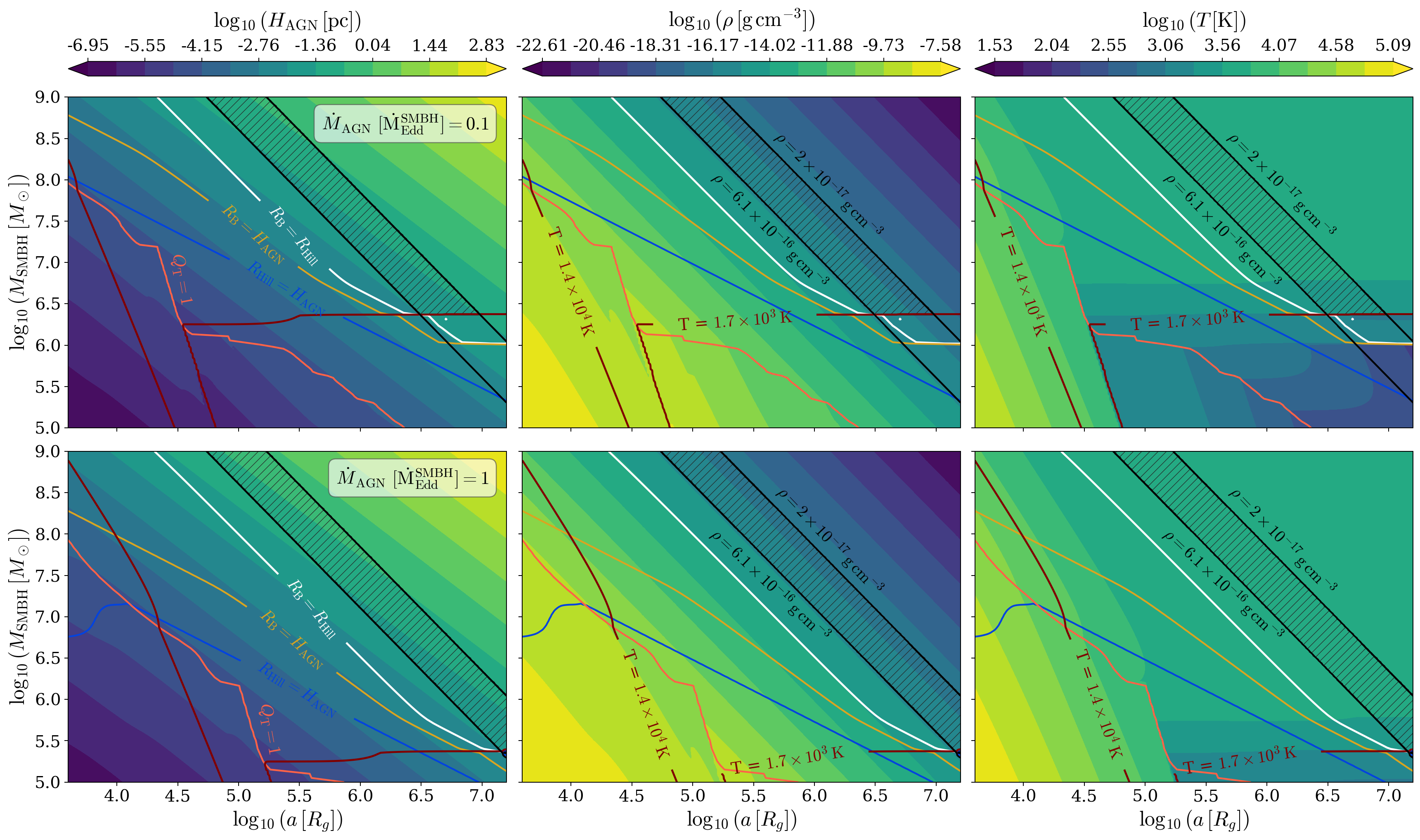}
    \caption{Disk scale height, mid-plane density, and temperature (left to right) predicted by the \citetalias{2003sirko} AGN disk model across the plane of central SMBH of mass $M_{\rm SMBH}$ (in $\rm{M}_\odot$) and orbital distance $a$ (in gravitational radii, $R_g \equiv GM_{\rm SMBH}/c^2$), for a disk viscosity $\alpha = 0.1$. The top row shows disk conditions for a constant mass flux $\dot{M}_{\rm AGN} = 0.1\,\dot{\rm M}_{\rm Edd}^{\rm SMBH}$ and the bottom row for $\dot{M}_{\rm AGN} = \dot{\rm M}_{\rm Edd}^{\rm SMBH}$. Overlaid contours mark the $Q_{\rm T} = 1$ boundary (orange), the locus where the Bondi radius equals the Hill radius, $R_{\rm B} = R_{\rm Hill}$ (white), where the Bondi radius equals the disk scale height, $R_{\rm B} = H_{\rm AGN}$ (yellow), and where the Hill radius equals the disk scale height, $R_{\rm Hill} = H_{\rm AGN}$ (blue). We also mark the minimum and maximum temperatures (dark red) and midplane densities (black) sampled by our simulation suite, $T = \left(1.7 \times 10^3~{\rm K},\ 1.4 \times 10^4~{\rm K}\right)$ and $\rho_0 = \left(2.04 \times 10^{-17}~{\rm g\,cm^{-3}},\ 6.12 \times 10^{-16}~{\rm g\,cm^{-3}}\right)$. The hatched region marks the portion of the $(a,\,M_{\rm SMBH})$ plane actually sampled by our simulation suite. The sampled $\rho$ range is deliberately narrow: the computational cost of simulations rises steeply with increasing $\rho$ and our results depend only weakly on $\rho$ as discussed in \S\ref{subsec:results_sim_suite}. Models were constructed using the \textsc{pAGN} code \citep{2024gangardt}.}
    \label{fig:AGN_params}
\end{figure*}

Our aim here is to evaluate conditions relevant to the AGN disk, specifically the outer regions of the disk, where embedded stars and compact objects are expected to form in situ, and where analytic models predict that feedback from the embedded objects themselves should suppress their accretion most strongly relative to the Bondi rate \citep{2022tagawa}. In order to simulate this environment, we first evaluate the relevant disk parameters predicted from analytic models and the accretion and outflow rates they imply. The structure of these regions is not well constrained. Standard $\alpha$-disk models become gravitationally unstable at large radii, where the Toomre parameter \citep{toomre1964}
\begin{equation}
    Q_{\rm T} = \frac{c_s \Omega}{\pi G \Sigma} = \frac{\Omega^2}{\sqrt{2} \pi G \rho}
\label{eq:Qtoomre}
\end{equation}
falls below unity, with $c_s$ the sound speed, $\Sigma$ the surface density, and $\Omega = \sqrt{G M_{\rm SMBH}/a^3}$ the Keplerian, orbital frequency a distance $a$ from an SMBH of mass $M_{\rm SMBH}$. The AGN disk models most commonly used counter this instability by invoking feedback from star formation, which supplies the pressure needed to hold the disk at marginal stability, $Q_{\rm T} \simeq 1$. Under this constraint, the mid-plane density is fixed at $\rho_0 = \Omega^2/\sqrt{2}\pi G Q_{\rm T}\propto M_{\rm SMBH}/a^3$. Still, $T$ and $H_{\rm AGN}$ vary with the chosen disk model, the most commonly invoked being those of \citet[hereafter \citetalias{2003sirko}]{2003sirko} and \citet[hereafter \citetalias{TQM05}]{TQM05}. In the former, the mass flux is held constant across the AGN disk, while the latter assumes the mass flux $\dot{M}_{\rm AGN}$ declines inward as star formation consumes gas in proportion to the heating required to support the disk. The reduced mass flux yields lower temperatures and scale heights than an \citetalias{2003sirko} model with the same outer-boundary conditions. While the \citetalias{TQM05} treatment is arguably more self-consistent, its inclusion of mass loss via star formation sets a maximum mass flux that can reach the disk interior, thereby narrowing the range of disk conditions it can realize. Moreover, both \citetalias{2003sirko} and \citetalias{TQM05} ignore the radiation pressure from accreting BHs, which has been shown to enhance temperature \citep{2024zhou} and scale height across the disk both when assumed to be the sole source of pressure support against gravitational collapse \citep{gilbaum2022} and when incorporated as a time-evolving population \citep{epstein-martin2025}. Since the specific mechanism maintaining marginal stability in these outer regions remains uncertain -- auxiliary support needs not even be thermal\footnote{Several works have shown that magnetic pressure support may be sufficient to prevent gravitational collapse in the outer disk \citep[e.g.][]{2024hopkins, 2026gerling-dunsmore}. An additional source of auxiliary pressure support may come from the very outflows from BHs we discuss here, in particular \citet{2026liu} show that shock interactions driven by mechanical feedback can enhance viscosity, thereby reducing the surface density in the outer disk.} -- we adopt the \citetalias{2003sirko} model, whose constant mass flux provides a single-parameter bracket of the local disk conditions.

Fig.\ref{fig:AGN_params} shows the disk scale height, mid-plane density, and temperature (left to right) predicted by the \citetalias{2003sirko} model as functions of $M_{\rm SMBH}$ and $a$, for a viscosity parameter $\alpha = 0.1$ and disk mass fluxes of $\dot{M}_{\rm AGN} = 0.1\,\dot{\rm M}_{\rm Edd}^{\rm SMBH}$ (top row) and $\dot{M}_{\rm AGN} = \dot{\rm M}_{\rm Edd}^{\rm SMBH}$ (bottom row), where
\begin{equation}
    \dot{\rm M}_{\rm Edd}^{\rm SMBH} = \frac{4\pi G M_{\rm SMBH}}{\eta\,\kappa_{\rm es}\,c},
\end{equation}
with $\eta = 0.1$ the radiative efficiency, $\kappa_{\rm es} \simeq 0.34~{\rm cm^2\,g^{-1}}$ the electron-scattering opacity for solar composition, and $c = 3\times 10^{10}\,\rm{cm \, s^{-1}}$ the speed of light. The maps were generated from an interpolated suite of these models computed with the \textsc{pAGN} code \citep{2024gangardt}.

We mark several contours that serve as useful reference points for the accretion and dynamics of an embedded BH. The white line traces where the Bondi and Hill radii are equal, $\rm R_{\rm B} = R_{\rm Hill}$. Here $\text{R}_{\rm B} = GM_{\rm BH}/c_s^2$ is the Bondi radius of the embedded BH \citep{bondi1952}, and $\text{R}_{\rm Hill} = a\,(M_{\rm BH}/3M_{\rm SMBH})^{1/3}$ is its Hill radius \citep{1999murray}. Our simulations are designed to mimic AGN conditions to the right of and above this line, where $\rm R_{\rm B} < R_{\rm Hill}$ and the accretion region is bounded by the Bondi radius; interior to it, the tidal field of the SMBH, rather than the gas sound speed, sets the size of the accretion region -- a background shear our uniform-medium setup does not capture. Note that in the $\dot{M}_{\rm AGN} = 0.1\,\dot{\rm M}_{\rm Edd}^{\rm SMBH}$ panels the cooler outer disk at low $M_{\rm SMBH}$ enlarges the Bondi radius, shrinking the region where $\rm R_{\rm B} < R_{\rm Hill}$ so that the white contour terminates at low SMBH mass.  The blue and yellow contours mark where the Bondi and Hill radii equal the disk scale height, $\text{R}_{\rm B} = H_{\rm AGN}$ and $\text{R}_{\rm Hill} = H_{\rm AGN}$. Where they exceed it, the accretion radius is larger than the disk thickness and the disk's vertical extent limits the available gas. 

Note that $\text{R}_{\rm B} < H_{\rm AGN}$ wherever $\rm R_{\rm B} < R_{\rm Hill}$, so that $H_{\rm AGN}$ has little to no impact on the Bondi accretion rate in the outer disk. This ordering is generic rather than a feature of the \citetalias{2003sirko} model. Writing the aspect ratio as $H_{\rm AGN}/a = c_s/(\Omega \, a)$, $R_{\rm B}$ becomes
\begin{equation}
    R_{\rm B} = \frac{G\,M_{\rm BH}}{c_s^2}= \frac{M_{\rm BH}}{M_{\rm SMBH}} \frac{a^3}{H_{\rm AGN}^2}, 
\end{equation}
so that the  relationship between $R_{\rm B}$, $H_{\rm AGN}$, and $R_{\rm Hill}$ reduce to powers of a single ratio, 
\begin{equation}
\begin{aligned}
        & \frac{R_{\rm B}}{H_{\rm AGN}} = 3 \left( \frac{R_{\rm Hill}}{H_{\rm AGN}}\right)^3, \\
        & \frac{R_{\rm B}}{R_{\rm Hill}} = 3 \left(\frac{R_{\rm Hill}}{H_{\rm AGN}}
\right)^{2}.
\end{aligned}  
\end{equation}
A Bondi-bounded accretion region, $R_{\rm B} < R_{\rm Hill}$, therefore requires $R_{\rm Hill}/H_{\rm AGN}< 3^{-1/2}$, which implies $R_{\rm B} < H_{\rm AGN}$. 

The orange contour marks the $Q_{\rm T} = 1$ boundary, which separates the two structural regimes of the disk. Interior to it the disk is gravitationally stable on its own and reduces to a standard $\alpha$-disk, whereas exterior to it the auxiliary pressure support invoked above is required to hold the disk at marginal stability. At the lower mass flux this boundary moves outward (compare the top and bottom rows of Fig.\ref{fig:AGN_params}): a less massive disk remains stable against its own self-gravity out to larger radii. The conditions we simulate lie entirely beyond this contour, in the outer, star-forming region where any in-situ population of embedded BHs is expected to reside.
\footnote{The reduced disk mass flux predicted by the \citetalias{TQM05} models pushes their $Q_{\rm T} = 1$ locus to still larger radii than in \citetalias{2003sirko} models with identical boundary conditions. Still, even in the \citetalias{TQM05} models, the densities sampled in this work correspond to marginally stable regions.}

Across much of the marginally stable ($Q_{\rm T} = 1$) outer disk the temperature is nearly independent of location, settling to $T \approx 4\times10^{3}$~K and falling below this value only in the lowest-mass systems, $M_{\rm SMBH}\lesssim 10^{6.5}\, \rm M_\odot$ ($10^{5.5} \, \rm M_\odot$) for $\dot{M}_{\rm AGN} = 0.1 \, \dot{\rm M}^{\rm SMBH}_{\rm Edd}$ ($1\,  \dot{\rm M}^{\rm SMBH}_{\rm Edd}$). At these temperatures the Rosseland opacity rises steeply with $T$ as H$^{-}$ becomes the dominant opacity source, and this steep $\kappa(T)$ dependence maintains the mid-plane temperature. Because the mid-plane density on the $Q_{\rm T}=1$ branch is a fixed function of $M_{\rm SMBH}$ and $a$, a higher mass flux instead manifests as a larger scale height, $H_{\rm AGN} \propto (\dot{M}_{\rm AGN}/\alpha)^{1/3}/\Omega$ \citep[see also the outer-disk scalings of][]{2024grishin}, visible in the leftmost panels of Fig.\ref{fig:AGN_params}. 

In this figure we also mark the specific conditions sampled by our simulations. The solid black lines bound the minimum and maximum ambient densities we test $2 \times 10^{-17} \leq \rho_0 \leq 6 \times 10^{-16}~{\rm g\,cm^{-3}}$, and the dark red lines the minimum and maximum ambient temperatures $1.7 \times 10^{3} \leq T \leq 1.4 \times 10^{4}~{\rm K}$ (see Table~\ref{table:sims} for simulation details). The hatched region marks the portion of the $(a,\,M_{\rm SMBH})$ plane our suite actually samples, where the disk model simultaneously falls within the labeled density bounds, exceeds our minimum sampled temperature, and satisfies $R_{\rm B} < R_{\rm Hill}$. Because the mid-plane density of any marginally stable disk is the same fixed function of $M_{\rm SMBH}$ and $a$ (Eq.~\ref{eq:Qtoomre}), these density bounds trace the same loci in the $(a,\,M_{\rm SMBH})$ plane across all models. Because the mid-plane density of any marginally stable disk is the same fixed function of $M_{\rm SMBH}$ and $a$ (Eq.~\ref{eq:Qtoomre}), the sampled densities correspond to the same locations in the $(M_{\rm SMBH}, a)$ plane across all models. The sampled temperatures span the full range expected across the outer disk, whereas our density sampling is more limited. We restrict the density range both because high-density runs are computationally expensive to evolve at adequate resolution and because, as our simulations progressed, we found the ambient density to have little effect on the evolution, as discussed in \S~\ref{subsec:results_sim_suite}. We note, however, that the densities we do sample fall in the low-opacity portion of the outer disk -- where the largest star-formation heating rates are required to maintain marginal stability, and hence where in-situ star formation and BH seeding are expected to peak -- so our suite targets the conditions most relevant for an embedded BH population.

% ----------------------------------------------------------------------------------------------------
% Sec: AGN_Params
% Subsec:super-Eddington_accretion
% ----------------------------------------------------------------------------------------------------

\subsection{Super-Eddington Accretion and Outflows}\label{subsec:super-Eddington_accretion}

\begin{figure*}
    \centering
    \includegraphics[width=0.98\linewidth]{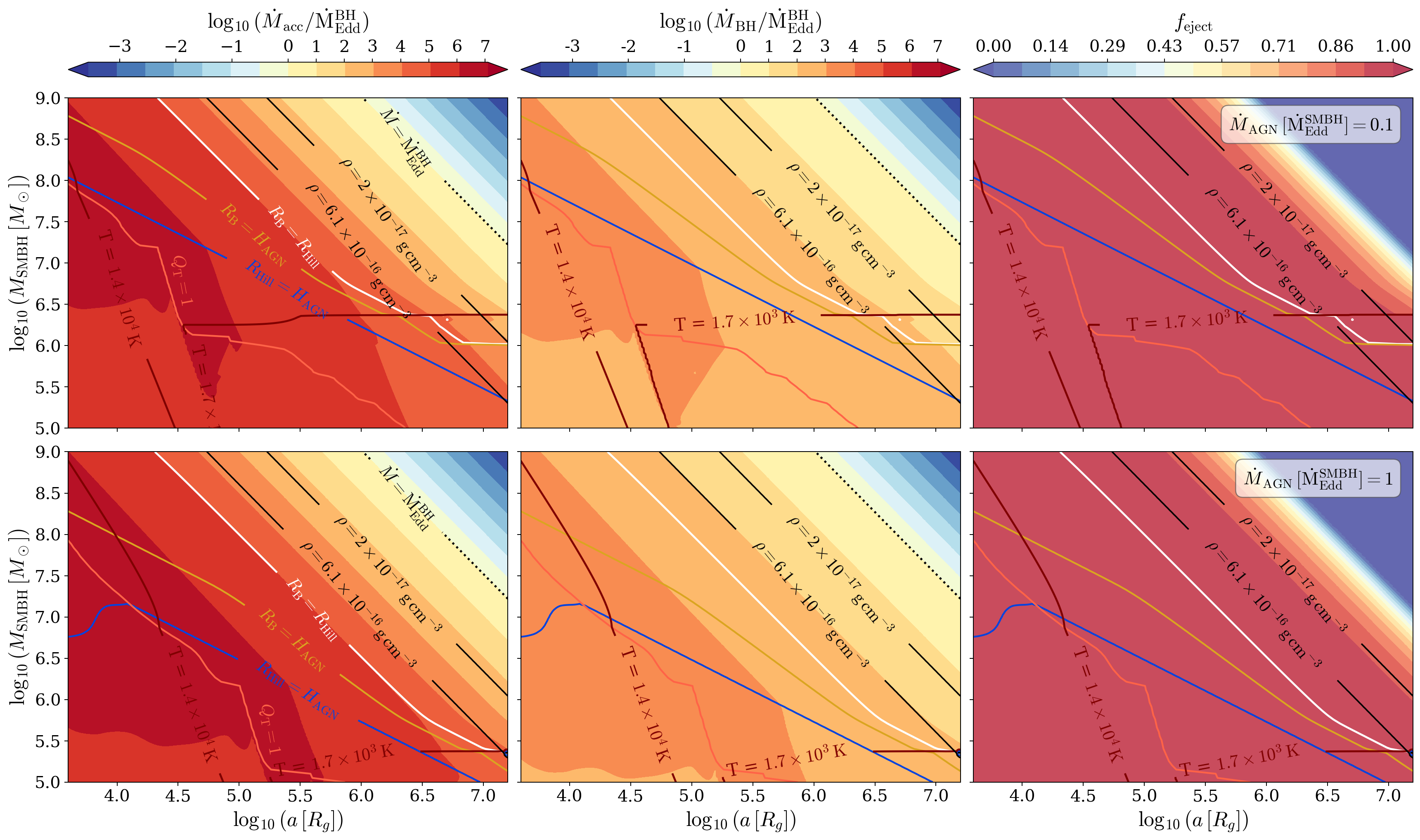}
    \caption{Accretion onto an embedded $10\,{\rm M}_\odot$ BH across the disk plane of central SMBH mass $M_{\rm SMBH}$ and orbital distance $a$, for the same disk models and parameters as in Fig.\ref{fig:AGN_params}. Left to right, the columns show the gas capture rate $\dot{M}_{\rm acc}$ (Eq.~\ref{eq:mdot_acc}), the growth rate $\dot{M}_{\rm BH}$ remaining after trapped-radiation-driven mass loss, and the ejected fraction $f_{\rm eject}$, computed with a trapping index $p = 0.5$ (Eq.~\ref{eq:mdotBH_trap}). The first two columns are plotted on a common logarithmic scale in units of the embedded BH's Eddington rate $\dot{\rm M}_{\rm Edd}^{\rm BH}$, with the color scale centered on $\dot{M} = \dot{\rm M}_{\rm Edd}^{\rm BH}$. The black dotted contour marks this Eddington locus in each panel; above and to the right of it, $\dot{M}_{\rm acc} < \dot{\rm M}_{\rm Edd}^{\rm BH}$ and, in the third column, $f_{\rm eject} = 0$. Overlaid as in Fig.\ref{fig:AGN_params} are the $Q_{\rm T} = 1$ boundary (orange), the loci $R_{\rm B} = R_{\rm Hill}$ (white), $R_{\rm B} = H_{\rm AGN}$ (yellow), and $R_{\rm Hill} = H_{\rm AGN}$ (blue), together with the minimum and maximum effective temperatures (dark red) and midplane densities (black) sampled by our simulation suite.}
    \label{fig:accretion_maps}
\end{figure*}

From the AGN disk parameters we can estimate the gas capture rate onto an embedded BH, which is set by the smaller of the accretion radii, $R_{\rm acc} = \min(R_{\rm B},R_{\rm Hill})$, the vertical extent of the disk relative to the capture radius, $\min\!\left(1,\,\frac{H_{\rm AGN}}{R_{\rm acc}}\right)$, and the relative gas velocity,  $v_{\rm eff} \simeq (c_s^2 + (\Omega R_{\rm Hill})^{2})^{1/2}$ \citep{2011kocsis, 2017stone, 2022tagawa},
\begin{equation}
    \dot{M}_{\rm acc} = e^{3/2}\,\pi\,\rho_0\,v_{\rm eff}\,R_{\rm acc}^2
    \min\!\left(1,\,\frac{H_{\rm AGN}}{R_{\rm acc}}\right),
\label{eq:mdot_acc}
\end{equation}
where $e\approx 2.718$ is Euler's number. Note that in the outer disk, where $R_{\rm B} < R_{\rm Hill}$ and $R_{\rm B}  <H_{\rm AGN}$, $\dot{M}_{\rm acc}$ reduces to the isothermal Bondi accretion rate,  $\dot{\rm M}_{\rm B} = e^{3/2}\pi\rho_0 (GM_{\rm BH})^2 c_s^{-3}$ \citep{bondi1952}. In units of the Eddington accretion rate this is, 
\begin{equation}
\begin{aligned}
    \frac{\dot{\rm M}_{\rm B}}{\dot{\rm M}_{\rm Edd}^{\rm BH}}
    &= \frac{e^{3/2}}{4}\,
       \frac{\eta\,\kappa_{\rm es}\,c\,G M_{\rm BH}\,\rho_0}{c_s^3}\\
    &\approx 55
    \left(\frac{\eta}{0.1}\right)
    \left(\frac{\kappa_{\rm es}}{0.34~{\rm cm^2\,g^{-1}}}\right)
    \left(\frac{M_{\rm BH}}{10\,{\rm M}_\odot}\right)
    \left(\frac{\mu}{1.22}\right)^{3/2}\\
    &\quad \times\left(\frac{\rho_0}{2\times10^{-17}~{\rm g\,cm^{-3}}}\right)
    \left(\frac{T_0}{10^4~{\rm K}}\right)^{-3/2}.
\end{aligned}
\label{eq:mdotB_edd}
\end{equation}
The first column of Fig.\ref{fig:accretion_maps} maps $\dot{M}_{\rm acc}$ in Eddington units across the disk plane. It rises inward, reaching $10^{3}\,\dot{\rm M}_{\rm Edd}^{\rm BH}$ near the $R_{\rm B} = R_{\rm Hill}$ locus, and remains above unity across most of the relevant parameter space (below and to the left of the dotted line) and across our simulation suite. 

Under these conditions the inflow is optically thick enough that radiation cannot diffuse outward faster than it is advected inward, and it becomes trapped within a characteristic radius  \citep{1979begelman, 1988abramowicz, 2005ohsuga}. Interior to $r_{\rm trap}$ the photon diffusion time exceeds the local inflow time, so radiation is carried inward with the gas rather than escaping. The trapping radius is
\begin{equation}
    r_{\rm trap} = \frac{3}{5}\,
    \frac{\dot{M}_{\rm acc}}{\dot{\rm M}_{\rm Edd}^{\rm BH}}\,r_{\rm min},
\label{eq:rtrap}
\end{equation}
where $r_{\rm min}$ is the inner edge of the circum-BH disk, taken to be the innermost stable circular orbit, $r_{\rm min} = 3R_{\rm s} = 6\,GM_{\rm BH}/c^2$, with $R_{\rm s}$ the Schwarzschild radius. Within the trapping region, radiation pressure drives strong outflows that strip mass from the inflow, so the flux that reaches and grows the BH falls below the capture rate as
\begin{equation}
    \dot{M}_{\rm BH} = \dot{M}_{\rm acc}
    \left(\frac{r_{\rm min}}{r_{\rm trap}}\right)^{p},
\label{eq:mdotBH_trap}
\end{equation}
with index $p \sim 0.5$--$0.7$ determined numerically \citep[e.g.][]{2022hu}. We adopt $p = 0.5$ throughout. The fraction of the captured inflow ejected before it can accrete is therefore
\begin{equation}
    f_{\rm eject} \equiv 1 - \frac{\dot{M}_{\rm BH}}{\dot{M}_{\rm acc}}
    = 1 - \left(\frac{5}{3}\,
    \frac{\dot{\rm M}_{\rm Edd}^{\rm BH}}{\dot{M}_{\rm acc}}\right)^{p},
\label{eq:feject}
\end{equation}
which we set to zero where $\dot{M}_{\rm acc} < \tfrac{5}{3}\,\dot{\rm M}_{\rm Edd}^{\rm BH}$.

The second column of Fig.\ref{fig:accretion_maps} shows the growth rate $\dot{M}_{\rm BH}$ that survives this mass loss. It rises inward as the capture rate does, reaching a maximum of order $10^{3}\,\dot{\rm M}_{\rm Edd}^{\rm BH}$ in the inner disk. Across the region our simulations sample, between the locus where accretion first becomes super-Eddington and the $R_{\rm B} = R_{\rm Hill}$ boundary, $\dot{M}_{\rm BH}$ is reduced by roughly an order of magnitude relative to $\dot{M}_{\rm acc}$ but remains at $10$--$100$ times the Eddington rate. In the third column, interior to the  $\dot{M}_{\rm acc} = \dot{\rm M}_{\rm Edd}^{\rm BH}$ locus, $f_{\rm eject}$ climbs steeply towards unity and the vast majority of encountered gas is ejected in the outflow.

Fig.\ref{fig:accretion_maps} illustrates three key points: First, that capture rates are super-Eddington across the vast majority of the disk, including the outer, Bondi-dominated region on which we focus. Second, they remain super-Eddington even after the outflow driven by trapped radiation is accounted for, with the growth rate saturating near $10^{3}\,\dot{\rm M}_{\rm Edd}^{\rm BH}$ in the inner disk. Third, the outflows themselves are correspondingly strong and heavily mass loaded, with the ejected flux approaching the gas capture rate wherever $f_{\rm eject} \to 1$. These estimates establish that feedback should matter throughout the disk, but they say nothing about how the ejected material interacts with the ambient medium once it leaves the trapping region. It is that interaction, and the additional suppression it imposes, to which we now turn.

% ----------------------------------------------------------------------------------------------------
% Sec: Theory
% ----------------------------------------------------------------------------------------------------

\section{Analytic description of jet-driven cocoons}
\label{sec:theory}

Previous analytical treatments describe jet-driven cocoons either as energy-conserving bubbles, whose hot interior retains the injected power, or as momentum-driven outflows, whose swept-up shell is pushed directly by the jet thrust. These limits provide useful reference points for our simulations, and we summarize them here. 

We organize the discussion around three ingredients. First (\S\ref{subsec:expansion_laws}) we review the idealized energy-conserving expansion laws for both spherical bubbles and elongated jet cocoons. Second (\S\ref{subsec:bondi_momentum}) we summarize the numerically obtained self-regulation condition of \citet{2021su, 2023su, 2025su} in this non-radiative regime. Third (\S\ref{subsec:momentum_driven}) we treat the opposing, momentum-driven limit, in which the interior has radiated its energy and the shell is driven by thrust alone. Departures of the simulations from these reference scalings then signal the physics the idealized laws omit -- time-dependent accretion, radiative cooling, turbulent mixing at the cocoon boundary, or the failure of the cocoon to remain a single pressure-confined structure.

% ----------------------------------------------------------------------------------------------------
% Sec: Theory
% Subsec:expansion_laws
% ----------------------------------------------------------------------------------------------------

\subsection{Energy-conserving expansion laws}
\label{subsec:expansion_laws}

We first summarize the energy-conserving expansion laws. Detailed derivations are given in Appendix~\ref{app:bubble_scalings}. Here we state only the governing momentum and energy equations and the resulting scalings.

For a spherical, pressure-driven bubble expanding into a homogeneous ambient medium of density $\rho_0$, the swept-up shell is accelerated by the pressure of the hot shocked wind interior,
\begin{equation}
    \frac{d}{dt}\left(M_{\rm sh}\dot R\right) = 4\pi R^2P_c,
\label{eq:theory_iso_momentum}
\end{equation}
where $M_{\rm sh}=4\pi\rho_0R^3/3$ is the mass of the swept up shell. In the absence of radiative losses, the interior energy evolves according to
\begin{equation}
    \frac{d}{dt} \left( \frac{P_cV_c}{\gamma-1} \right) = \dot E_{\rm in}  - P_c\frac{dV_c}{dt},
\label{eq:theory_iso_energy}
\end{equation}
where $\gamma = 5/3$, $V=4\pi R^3/3$ and
\begin{equation}
    \dot E_{\rm in} = \epsilon\dot E_j = \frac{\epsilon}{2}\dot M_j v_j^2. 
\end{equation}
Here $\epsilon$ is the fraction of the jet power retained by the cocoon and measures how much of the shocked jet energy is thermalized in the cocoon interior. For a strong adiabatic shock the post-shock flow retains the bulk of the injected energy, and we adopt $\epsilon\simeq15/16$ \citep{2019el-badry}. 

For constant $\dot E_{\rm in}$, Eqs.~\eqref{eq:theory_iso_momentum}--\eqref{eq:theory_iso_energy} give the classical energy-conserving bubble solution \citep{1977weaver}
\begin{align}
    R_{\rm sph}(t)
    &=
    \left(\frac{125}{154\pi}\right)^{1/5} \left(\frac{\dot E_{\rm in}}{\rho_0}\right)^{1/5} t^{3/5},
    \label{eq:theory_R_sph}
    \\
    P_{\rm sph}(t)
    &=
    \frac{5}{22\pi} \left(\frac{125}{154\pi}\right)^{-3/5} \dot E_{\rm in}^{2/5}\rho_0^{3/5}t^{-4/5}.
    \label{eq:theory_P_sph}
\end{align}

For an anisotropic jet-driven cocoon, the same conservation laws apply, but the axial and lateral directions are controlled by momentum and energy conservation respectively. We approximate the bipolar cocoon as a cylinder of lateral radius $R$ and one-sided height $z$, with volume $V_c\simeq2\pi R^2z$. The lateral expansion is pressure-driven, while the jet head advances by balancing jet thrust against ambient ram pressure,
\begin{equation}
\rho_0 A_h\dot z^2
\simeq
\frac{\dot\Pi_j}{2},
\label{eq:theory_head_balance}
\end{equation}
where $\dot\Pi_j=\dot M_jv_j$ is the total bipolar jet thrust. If the effective head area is set by the cocoon width, $A_h\simeq\pi R^2$, the energy-conserving elongated-cocoon solution is \citep{1989begelman, 2011bromberg}
\begin{align}
    R(t)
    &=
    A t^{2/3}
    \propto
    \epsilon^{1/3}\dot M_j^{1/6}v_j^{1/2}\rho_0^{-1/6}t^{2/3},
    \label{eq:theory_R_aniso}
    \\
    z(t)
    &=
    \frac{3K}{A}t^{1/3}
    \propto
    \epsilon^{-1/3}\dot M_j^{1/3}\rho_0^{-1/3}t^{1/3},
    \label{eq:theory_z_aniso}
    \\
    P_c(t)
    &=
    \frac{4}{9}\rho_0A^2t^{-2/3}
    \propto
    \epsilon^{2/3}\dot M_j^{1/3}v_j\rho_0^{2/3}t^{-2/3}.
    \label{eq:theory_P_aniso}
\end{align}
where
\begin{equation}
    K\equiv
    \left(\frac{\dot\Pi_j}{2\pi\rho_0}\right)^{1/2},
    \qquad
    A\equiv
    \left[
    \frac{9}{68}
    \frac{\dot E_{\rm in}}{\pi\rho_0K}
    \right]^{1/3}.
\label{eq:KA}
\end{equation}

In order to apply these solutions in a physically motivated way, it is worth stating precisely the geometry of the outflow expected in the embedded BH/AGN context and how we model it. The mechanism we primarily have in mind is the radiation-pressure-driven outflow from a super-Eddington CBD. Radiation-hydrodynamic simulations of these flows find the ejecta to be strongly anisotropic. In the models of \citet{2021kitaki} the mechanical energy flux peaks at $\sim 15^\circ$  from the rotation axis, within an evacuated funnel, while the bulk of the mass flux emerges near $\sim 80^\circ$, close to the disk plane.  Our injection idealizes the funnel component of such a flow and pushes it to the extreme-collimation limit, adopting an outflow opening angle $\theta_j = 1^\circ$ and neglecting the angular separation between the energy- and mass-dominated components found in \citet{2021kitaki}.  This setup maximizes the contrast between axial and lateral energy deposition, providing the most conservative test of whether anisotropic feedback can regulate accretion, since a wider outflow interrupts the inflow more readily. 

Another possibility is that an embedded BH drives a relativistic jet, launched magnetically from the BH spin \citep{1977blandford} or from the inner disk \citep{1982blandford_payne}. The jet velocities we inject, $v_j = 10^3$--$10^4~{\rm km\,s^{-1}} \lesssim 0.03c$, are firmly sub-relativistic, but our methodology may still be relevant to a relativistic jet if it is mass-loaded and decelerated by entrainment. Relativistic hydrodynamic simulations find that instabilities can grow at the jet-environment contact discontinuity and mix external gas \citep{2007perucho, 2018gourgouliatos, 2023abolmasov, 2024costa}. The entrained material carries off the jet's momentum, in some cases slowing it to sub-relativistic speeds \citep{2020rossi}. Whether this carries over to a magnetically launched jet is uncertain, since even weak fields have been shown to suppress the relevant local instabilities \citep{2020gottlieb}. In either case our results are not tied to the particular launching mechanism, although our modeled setup is primarily motivated by anisotropic outflows expected from super-Eddington winds. 

For brevity we refer throughout to the ``jet'' and to jet-driven cocoons, following the standard nomenclature of the jet feedback mechanism and of the analytic cocoon literature we build on \citep{1989begelman}.  This usage should be read as shorthand for a collimated, mass-loaded outflow of the kind described above and does not imply a magnetic launching mechanism. 

Under this assumption of a `jet-like' outflow with a narrow opening angle, cocoons begin in an elongated state described by Eqs.~\eqref{eq:theory_R_aniso}--\eqref{eq:theory_P_aniso}. However, because the lateral radius grows as $R\propto t^{2/3}$ against $z\propto t^{1/3}$ for the head, the aspect ratio $z/R$ falls steadily towards unity. Once $z\simeq R$ the cocoon isotropizes and evolves thereafter as an approximate spherical bubble [Eqs.~\eqref{eq:theory_R_sph}--\eqref{eq:theory_P_sph}]. 

Setting $z=R$ in the elongated solution gives the isotropization radius $R_{\rm iso}=9K^2/A^3$, or, with $K$ and $A$ from Eq.~(\ref{eq:KA}),
\begin{equation}
    R_{\rm iso} = \frac{68\pi\rho_0 K^3}{\dot E_{\rm in}}
    = \frac{68}{\sqrt{2\pi}}\,\frac{1}{\epsilon}
    \left(\frac{\dot M_j}{\rho_0 v_j}\right)^{1/2}.
\label{eq:Riso}
\end{equation}
Which of the two laws applies at a given radius is therefore set by $R_{\rm iso}$.  We return to this comparison in \S\ref{subsec:bondi_momentum} once the regulated jet mass flux has been determined.

% ----------------------------------------------------------------------------------------------------
% Sec: Theory
% Subsec:bondi_momentum
% ----------------------------------------------------------------------------------------------------

\subsection{Bondi-radius momentum regulation}
\label{subsec:bondi_momentum}

The energy-conserving expansion laws above describe how a cocoon propagates for a prescribed jet power. Physically and in simulations, however, the jet power is coupled to the accretion rate, so the cocoon and inflow are interdependent quantities. \citet{2023su} found that the relationship between inflow and outflow is well described by a momentum-regulated picture in which the jet mass flux adjusts until the outward momentum flux of the cocoon balances the inward momentum flux of the gas feeding the compact object.

The natural scale for this balance is the Bondi radius, where the inflow velocity is of order the ambient sound speed, $u(R_{\rm B})\sim c_s$. The inward momentum flux through the Bondi sphere is the Bondi mass flux carried by the inflow or
\begin{equation}
    \dot\Pi_{\rm in, B} \simeq e^{3/2}\pi R_{\rm B}^2\rho_0 c_s^2 .
\label{eq:bondi_momentum_flux}
\end{equation}

The outward, cocoon-driven momentum flux, is determined by the shape of the shock front as it crosses the Bondi radius. Following \citep{2023su}, we characterize it by an isotropic expansion velocity  $V_{\rm iso,B}$, defined so that the cocoon delivers the same total momentum flux as a spherical shock expanding at that speed,
\begin{equation}
    \dot\Pi_{\rm out,B} \simeq 4\pi R_{\rm B}^2\rho_0 V_{\rm iso,B}^2 .
\label{eq:cocoon_momentum_flux}
\end{equation}
If $R_{\rm iso} < R_{\rm B}$ the cocoon has already isotropized, the shock front is spherical, and $V_{\rm iso,B}$ is simply the bubble's radial velocity at $R_{\rm B}$. If $R_{\rm iso} > R_{\rm B}$ the cocoon is still elongated and the momentum flux carries both an axial and a lateral component. In principle $\dot\Pi_{\rm out,B}$ should then be evaluated as the scalar surface integral of the normal ram pressure over the cocoon boundary. However, because the jet head crosses the Bondi surface before the lateral edge does, it is the lateral surface that intercepts the inward momentum flux, and we estimate $V_{\rm iso,B}$ from the lateral expansion velocity alone. The two cases give 
\begin{equation}
    V_{\rm iso,B} \simeq
    \begin{cases}
    \dot R_{\rm sph}(R_{\rm B}),
    & R_{\rm iso} < R_{\rm B},\\[2.4ex]
    \sqrt{2}\,\dot R(R_{\rm B}),
    & R_{\rm iso} > R_{\rm B}.
\end{cases}
\label{eq:viso_def}
\end{equation}
The factor of two in the momentum flux is fixed by requiring that the estimate recover the full outward flux  $V_{\rm iso,B}^2\approx\dot z^2+\dot R^2$, when the lateral and axial contributions are equal.

The regulation condition $\dot\Pi_{\rm out,B}\sim\dot\Pi_{\rm in,B}$ is then equivalent to
\begin{equation}
    V_{\rm iso,B} \simeq c_s .
\label{eq:viso_regulation}
\end{equation}
Because the post-shock cocoon pressure scales as $P_c\propto\rho_0 V_{\rm iso,B}^2$, this may also be written as a pressure-balance condition,
\begin{equation}
    P_c(R_{\rm B})\propto \rho_0 c_s^2 .
\label{eq:pressure_regulation}
\end{equation}
If the cocoon pressure at $R_{\rm B}$ exceeds the ram pressure of the inflow, the cocoon can displace the gas feeding the BH, suppressing accretion. The reduced accretion rate then lowers the jet power, while expansion and interface losses decrease the cocoon pressure. Conversely, if the cocoon pressure falls below the Bondi ram pressure, gas can refill the central region, increasing the accretion rate. Thus the Bondi radius is the scale at which the outflow first competes directly with the gravitationally focused inflow.

Inserting the expansion laws of \S\ref{subsec:expansion_laws} into the regulation condition in Equation \eqref{eq:viso_regulation} fixes the regulated jet mass flux,
\begin{equation}
    \frac{\dot M_j}{\dot M_{\rm B}} \simeq
    \begin{cases}
    \dfrac{308}{27\,e^{3/2}}\,\epsilon^{-1}\left(\dfrac{c_s}{v_j}\right)^{2},
    & R_{\rm iso} < R_{\rm B},\\[2.4ex]
    \dfrac{289}{2\,e^{3/2}}\,\epsilon^{-2}\left(\dfrac{c_s}{v_j}\right)^{3},
    & R_{\rm iso} > R_{\rm B},
\end{cases}
\label{eq:mdot_regulated}
\end{equation}
where the upper and lower branches define the isotropic- and cylindrical-adiabatic regulated fluxes $\dot M_{\rm iso,ad}$ and $\dot M_{\rm cyl,ad}$ used in Figs.\ref{fig:NO_COOL_TS}, \ref{fig:cool_ts},  and \ref{fig:suite_grid}.

The jet velocity is the only remaining parameter fixed neither by the AGN environment nor by the momentum balance, and in this framework can be used to determine whether the cocoon is spherical or elongated when it crosses $R_{\rm B}$. Re-writing $R_{\rm iso}$ (Eq.~\eqref{eq:Riso}) in Bondi units yields
\begin{equation}
\frac{R_{\rm iso}}{R_{\rm B}} = 34\sqrt{2}\,\frac{e^{3/4}}{\epsilon}
\left(\frac{\dot M_j}{\dot M_{\rm B}}\right)^{1/2}
\left(\frac{c_s}{v_j}\right)^{1/2}.
\label{eq:Riso_RB}
\end{equation}
Closing this relation with Eq.~\eqref{eq:mdot_regulated}, the elongated branch gives $R_{\rm iso}/R_{\rm B} = 578\,\epsilon^{-2}(c_s/v_j)^2$, the factors of $e$ cancelling exactly, so the cocoon isotropizes within the Bondi radius for $v_j/c_s \gtrsim \sqrt{578}/\epsilon \simeq 26$. The isotropic branch gives $R_{\rm iso}/R_{\rm B} = 34\sqrt{616/27}\, \epsilon^{-3/2}(c_s/v_j)^{3/2}$ and hence $v_j/c_s\gtrsim32$, the same condition to order unity. The offset between the two reflects the differing normalizations of the spherical and elongated laws, so the transition is best read as a band, $v_j/c_s\sim25$--$30$. Across our suite $v_j/c_s$ runs from $\sim3\times10^2$ to $\sim9\times10^3$, more than an order of magnitude above this threshold.

Taken at face value this places every run in the spherical branch, and does so for embedded BHs generally, since $c_s\sim$ a few $\rm km\,s^{-1}$ in the outer disk while any plausible jet velocity is orders of magnitude larger. However, this inference holds only within the adiabatic parameterization from which $R_{\rm iso}$ is derived. Any process that drains that energy, whether radiative losses at the cocoon surface or turbulent mixing across the contact discontinuity, weakens the lateral expansion while leaving the thrust-driven advance of the head untouched, and so defers isotropization to larger radii. The adiabatic estimate should therefore be read as a lower bound on $R_{\rm iso}$. We return to this issue in \S\ref{subsec:results_sim_suite}, where we find that the simulated cocoons retain a modest but persistent elongation across the full suite despite sitting far above the nominal threshold.

% ----------------------------------------------------------------------------------------------------
% Sec: Theory
% Subsec:momentum_driven
% ----------------------------------------------------------------------------------------------------

\subsection{Momentum-driven cocoons and the radiative limit}
\label{subsec:momentum_driven}

The expansion laws of \S\ref{subsec:expansion_laws} assume the cocoon interior retains injected jet energy. In the opposite, radiative limit the shocked gas radiates its thermal energy faster than it can do work, so the interior pressure drops out and the swept-up shell is driven directly by the injected momentum flux, or thrust, $\dot\Pi_j = \dot M_jv_j$.

We treat this limit as spherical by construction. A cocoon is elongated because its over-pressured interior inflates the sidewall while the head advances under thrust. Once that pressure has been radiated away no lateral cocoon forms at all and what remains is a bare jet head with oblique shocks along the beam. In this regime an anisotropic outflow's ram pressure greatly exceeds that of the ambient gas, $\rho_j v_j^2\gg\rho_0 c_s^2$, so that the ejecta is strongly under-confined and spreads laterally as it advances rather than remaining within the injection cone \citep{2023chen}. We therefore treat the fully radiative, momentum-driven shell as approximately spherical.

For a spherical shell of mass $M_{\rm sh}=\tfrac{4}{3}\pi\rho_0 R^3$ the momentum equation is
\begin{equation}
    \frac{\mathrm{d}}{\mathrm{d}t}\!\left(M_{\rm sh}\dot R\right)=\dot M_j v_j ,
\label{eq:mom_driven_eom}
\end{equation}
which for a constant thrust integrates to
\begin{equation}
    R(t)=\left(\frac{3\dot M_j v_j}{2\pi\rho_0}\right)^{1/4}t^{1/2}.
\label{eq:mom_driven_R}
\end{equation}
This expansion is shallower than the energy-conserving $R\propto t^{3/5}$ (Eq. ~\eqref{eq:theory_R_sph}) because only the delivered momentum drives the shell, with no contribution from the $P\,\mathrm{d}V$ work of a confined hot interior. 

Imposing the regulation condition of \S\ref{subsec:bondi_momentum}, $V_{\rm iso, B}=c_s$, on the derivative of Eq.~\eqref{eq:mom_driven_R} gives the momentum-driven regulated flux
\begin{equation}
    \frac{\dot M_j}{\dot M_{\rm B}}=\frac{8}{3} e^{-3/2}\,\frac{c_s}{v_j}
    \;\simeq\;0.6\,\frac{c_s}{v_j}.
\label{eq:mom_driven_regulated}
\end{equation}
This is linear in $c_s/v_j$, in contrast to the adiabatic $(c_s/v_j)^2$ of Eq.~\eqref{eq:mdot_regulated}. It is the momentum-regulated reference plotted in Figs.~\ref{fig:NO_COOL_TS}, \ref{fig:cool_ts}, and \ref{fig:suite_grid} below, and it bounds the regulated rate from above.

We do not, however, expect our cocoons to reach this fully radiative limit. The relevant test is whether the shocked gas can cool within a crossing time, $t_{\rm cross}=R/v_j$. The hottest and slowest-cooling component is the post-shock jet material, with a strong post-shock temperature
\begin{equation}
    T_{\rm ps}=\frac{3}{16}\frac{\mu m_p}{k_B}\,v_j^2
    \approx1.4\times10^{9}
    \left(\frac{v_j}{10^{4}\,\mathrm{km\,s^{-1}}}\right)^{2}\,\mathrm{K},
\label{eq:Tps}
\end{equation}
with mean molecular weight $\mu=0.62$ for the ionized post-shock gas. At these temperatures cooling is dominated by thermal bremsstrahlung, $\Lambda_{\rm ff}(T)\approx1.7\times10^{-27}\,T^{1/2} \,\mathrm{erg\,cm^{3}\,s^{-1}}$, which is $\approx6\times10^{-23}\,\mathrm{erg\,cm^{3}\,s^{-1}}$ at $T_{\rm ps}$ and essentially independent of metallicity. The corresponding cooling time, evaluated even at the ambient density $\rho_0$ -- an overestimate of the density of the cocoon interior and hence a lower bound on $t_{\rm cool}$ -- is
\begin{equation}
\begin{aligned}
    t_{\rm cool} &=\frac{3k_BT_{\rm ps}\,m_p}{X_H\rho_0\Lambda_{\rm ff}(T_{\rm ps})}\\
    &\approx30
    \left(\frac{v_j}{10^{4}\,\mathrm{km\,s^{-1}}}\right)
    \left(\frac{\rho_0}{2\times10^{-17}\,\mathrm{g\,cm^{-3}}}\right)^{-1}
    \,\mathrm{yr}.
\end{aligned}
\label{eq:tcool_ps}
\end{equation}
For the post-shock gas to radiate within a crossing time the cocoon would need to reach
\begin{equation}
\begin{aligned}
    R_{\rm rad}\equiv v_j \,t_{\rm cool}
    \approx300
    \left(\frac{v_j}{10^{4}\,\mathrm{km\,s^{-1}}}\right)^{2}
    \left(\frac{\rho_0}{2\times10^{-17}\,\mathrm{g\,cm^{-3}}}\right)^{-1}
    \,\mathrm{mpc},
\end{aligned}
\label{eq:Rrad}
\end{equation}
or approximately $90 \, R_{\rm B}$ against the Bondi radius $R_{\rm B}=GM_{\rm BH}\mu_0m_p/k_BT_{\rm 0}\approx3.7\,
(M_{\rm BH}/10\,M_\odot)(T_{0}/1741\,\mathrm{K})^{-1}$\,mpc,  with
$\mu_0=1.22$ for the neutral ambient medium. 

Radiative relaxation of the hot interior therefore requires the cocoon to reach at least the Bondi radius, and one to two orders of magnitude beyond it under fiducial conditions. Our cocoons stall near $R_{\rm B}$ in every run (\S\ref{subsec:results_sim_suite}), well short of $R_{\rm rad}$ across most of our suite. The one exception is the slowest jet sampled, $v_j=10^{3}\,\mathrm{km\,s^{-1}}$, for which the steep scaling $R_{\rm rad}\propto v_j^2$ brings $R_{\rm rad}$ down to $\simeq R_{\rm B}$. The post-shock gas in this case may cool within a crossing time, pushing these runs closer to the momentum-driven limit than the rest. The momentum-driven solution and its $\propto c_s/v_j$ regulated flux thus represent an upper bound.

% ----------------------------------------------------------------------------------------------------
% Sec:methods
% ----------------------------------------------------------------------------------------------------

\section{Methods: numerical simulations of jet-driven cocoons}\label{sec:methods}

To study the evolution of super-Eddington, outflow-driven cocoons, we simulate a uniform box of gas subject to feedback from a central $10\,\rm M_\odot$ BH, with ambient conditions chosen to mimic those of an AGN disk. Our three-dimensional hydrodynamic simulations use \textsc{GIZMO} \citep{2015hopkins} and its meshless finite-mass (MFM) hydrodynamics solver together with the FIRE-2 cooling and heating implementation, which includes photoelectric and photoionization heating, as well as optically thin cooling from collisional, Compton, fine-structure, recombination, atomic, and molecular processes \citep{2018hopkins}.

Our numerical setup is similar to that employed in \citet{2023su} and \citet{2025su}, who studied jet feedback from a $100\,\rm M_\odot$ seed BH in low-metallicity ($Z \sim 10^{-4}\,Z_\odot$) gas representative of high-redshift protogalaxies. Our adopted jet model and resolution refinement scheme follow the same algorithmic setup, described in more detail in the subsections below. However, our models invoke much higher gas densities ($n_0 \geq 10^7\,\mathrm{cm}^{-3}$) and test both low ($Z = 10^{-5}\,Z_\odot$) and high metallicities ($Z = Z_\odot$) to approximate the physical conditions expected in an AGN disk. Of these, solar metallicity is representative of a typical AGN disk, whose gas is generally enriched \citep{1999hamann, 2002hamann, 2006nagao, 2022wang}, whereas the extremely low-metallicity case is not expected except perhaps at early times or within small, poorly mixed regions of the disk \citep{2011husemann}. We include the latter primarily as an academic limit that bridges the low-metallicity protogalactic regime of \citet{2023su, 2025su} to the enriched AGN context. The combination of high density and solar metallicity substantially enhances the radiative cooling rate and places our simulations in a more strongly cooled regime than those of \citet{2023su, 2025su}. 

Since the high-density ambient medium in our simulations would otherwise be susceptible to Jeans fragmentation, we neglect gas self-gravity. This omission has little effect on the accretion and feedback we model. For our fiducial conditions ($n_0 = 10^7\,\mathrm{cm}^{-3}$, $T_0 = 3\times10^3\,\rm K$) the gas mass enclosed within the Bondi radius is $M_{\rm gas}(<R_{\rm B}) \approx 0.012\,\rm M_\odot$, about $10^{-3}$ of the $10\,\rm M_\odot$ BH, so the gas contributes negligibly to the potential, and it stays below a few per cent of the BH mass even at the highest densities we simulate. On the larger scale of the disk, gas at these densities is not expected to collapse freely either, being held near marginal stability by the background Keplerian shear and by the auxiliary pressure support that the cumulative feedback from star formation and accretion onto embedded compact objects provides (as discussed in \S\ref{sec:AGN_params}).

% ----------------------------------------------------------------------------------------------------
% Sec:methods
% subsec:setup_init
% ----------------------------------------------------------------------------------------------------

\subsection{Initial Conditions}\label{subsec:setup_init}

The ambient medium is initialized with uniformly distributed gas particles at constant density $\rho_0$, metallicity $Z_0$, and specific internal energy $u_0$. We set the specific internal energy $u_0$ such that it yields a temperature $T_0 = (\gamma - 1)\,u_0\,\mu  m_p / k_B$, where $m_p = 1.67\times10^{-24}\,\mathrm{g}$ is the proton mass, $k_B = 1.38\times10^{-16}\,\mathrm{erg\,K^{-1}}$ is Boltzmann's constant, $\gamma = 5/3$ is the adiabatic index, and the mean molecular weight is that of neutral atomic gas $\mu_{\rm atomic} = 4/(1+3X) = 1.22$ for hydrogen mass fraction $X = 0.76$. The nominal $T_0$ values listed in Table~\ref{table:sims} are therefore the temperatures at which the gas is
initialized. 

We impose a temperature floor ($T_{\rm floor}$) across all simulations, except in our fiducial fid.no\_cool model. Without it, radiative losses would rapidly drive most of the gas down to $\sim 10\,\mathrm{K}$ regardless of the initial thermal setup. Because we do not explicitly model the radiative heating that would offset cooling in a real disk, the floor stands in for a state in which cooling is balanced by heating. \textsc{gizmo} converts $T_{\rm floor}$ internally using the specific energy floor $u_0$, but assuming fully molecular gas whenever low-temperature cooling is enabled, $\mu_{\rm mol} = [X/2 + (1-X)/4 + 1/28]^{-1} = 2.10$. Since the gas that actually settles to the floor is neutral atomic, the realized floor temperature is lower than $T_0$ by the ratio of the two mean molecular weights: $T_{\rm floor} = T_0\,(\mu_{\rm atomic}/\mu_{\rm mol}) \approx 0.58\,T_0$. For our fiducial cooling runs this is $\approx 1741\,\mathrm{K}$ rather than the nominal $3000\,\mathrm{K}$. 

The adiabatic run (fid.no\_cool in Table~\ref{table:sims}) is the exception. With cooling disabled the gas never leaves its initial thermal state. The ambient temperature in this run is the nominal $T_0 = 3000$\,K, and the corresponding sound speed, Bondi radius and Bondi crossing time $\tau_B = R_{\rm B}/c_s$ are  $c_s = 4.5\,\mathrm{km\,s^{-1}}$, $R_{\rm B} = 2.12$\,mpc and $\tau_{\rm B} = 460$\,yr, against $3.4\,\mathrm{km\,s^{-1}}$, $3.65$\,mpc and $1040$\,yr in the radiative runs at the same nominal temperature.

Gas resolution increases towards the jet axis following the super-Lagrangian mesh refinement scheme implemented in \citet{2021su} and replicated in subsequent papers, with the target particle mass scaling linearly with cylindrical radius $\propto R_{\rm cyl} = \sqrt{x^2 + y^2}$ down to a minimum target mass resolution, $m_{\rm g,min}$ inside $R_{\rm cyl} < r_{\rm min}$, 
\begin{equation}
    m_{\rm g}(R_{\rm cyl}) = 10^{-6} \, {\rm M_\odot} \, \left(\frac{m_{\rm g,min}}{10^{-6} \, {\rm M_\odot}}\right) \max\!\left(\frac{R_{\rm cyl}}{r_{\rm min}},\;1\right)
\label{eq:mass_resolution}
\end{equation}
where $r_{\rm min} = L_{\rm Box}/32$ and $L_{\rm box}$ is the simulation box length. Gas resolution elements are automatically merged or split appropriately to maintain this mass resolution at all times. For the values adopted in most of our
runs, $m_{\rm g,min} = 10^{-6}\,{\rm M_\odot}$ and $L_{\rm box} = 100\times10^{-3}\,{\rm pc}$,
the fully refined core extends to $r_{\rm min} = 3.1\times10^{-3}\,{\rm pc}$ and the target
mass rises to $16\,m_{\rm g,min} = 1.6\times10^{-5}\,{\rm M_\odot}$ at the box edge. This hierarchical set-up allows for extremely high resolution around the $z$-axis where the jet is launched. Moreover, because $R_{\rm B} = 2.12\times10^{-3}\,{\rm pc}$ lies inside $r_{\rm min}$ under fiducial conditions, the Bondi sphere sits inside the uniformly refined core and is resolved by $N_{\rm g}(<R_{\rm B})\approx 1.2\times10^{4}$ resolution elements. 

% ----------------------------------------------------------------------------------------------------
% sec:methods
% subsec:setup_accretion
% ----------------------------------------------------------------------------------------------------

\subsection{Black Hole Accretion and Jet Spawning Algorithm}\label{subsec:setup_accretion}

Our simulations follow the gravitational capture of gas particles directly \citep{2016hopkins, 2021angles-alcazar}, implementing their subsequent accretion onto the BH via a subgrid $\alpha$-disk prescription. Candidate particles are drawn from the BH's neighbour kernel, whose radius is solved dynamically each timestep and capped at a maximum search radius of $2\times10^{-3}$~pc.\footnote{The effective neighbour number of the BH kernel is the standard value for our MFM solver ($N_{\rm ngb}=32$) multiplied by a BH-specific enhancement factor of 3. This enhancement prevents noisy estimates of environmental quantities (density, relative velocity, and angular momentum) from entering the accretion and feedback routines.} A candidate is accreted if it is gravitationally bound to the BH and its estimated apocentric radius lies within the sink radius $r_{\rm acc}$. The sink radius is tied to the BH's gravitational softening length $\epsilon_{\rm BH}$ as $r_{\rm acc} \approx 50\,\epsilon_{\rm BH}$, which for our fiducial resolution gives $r_{\rm acc} \approx 1.3\times10^{-4}$~pc.

Once captured, the particle mass is added to the subgrid $\alpha$-disk reservoir of total mass $M_\alpha$, which drains onto the BH at the rate $\dot{M}_{\rm acc} = M_\alpha/t_{\rm disk}$. The draining time $t_{\rm disk}$ is a parameter of the subgrid model, its adopted value motivated by the viscous timescale of a \citet{1973shakura} disk, $t_{\rm disk} \simeq t_\nu = r^2 \Omega / (\alpha c_s^2)$.\footnote{This estimate assumes a standard thin $\alpha$-disk \citep{1973shakura}, but the structure of the unresolved CBD is not well determined. In the strongly super-Eddington regime expected for embedded BHs, the disk is more likely radiatively inefficient and geometrically thick---a slim, advection-dominated disk \citep{1988abramowicz, 2005ohsuga}---for which the draining time would instead be of order $\alpha^{-1}\Omega^{-1}$, considerably shorter than the thin-disk estimate. This sensitivity to the assumed prescription is part of why we vary $t_{\rm disk}$ over a broad range and confirm that our conclusions do not depend on it (Appendix~\ref{app:t_disk_res}).} Taking $r$ to be the sink radius $r_{\rm acc} \sim 10^{-4}$~pc, a viscosity parameter $\alpha = 0.1$, a Keplerian orbital frequency $\Omega_{\rm BH} = \sqrt{G M_{\rm BH}/r^3}$ evaluated at that radius, and $c_s \sim 5$~km/s yields $t_{\rm disk} \sim 10^3$~yr. We therefore adopt a fiducial $t_{\rm disk} = 10^3$~yr, holding it fixed across the suite for consistency despite the varied draining time expected across simulated environmental parameters. We do not expect this choice to significantly alter our main conclusions. \citet{2023su} found the jet and cocoon dynamics to be largely insensitive to the draining time, with the regulated accretion rate varying by only a factor of $\sim 2$ across $t_{\rm disk} \in \{100, 1000, 10000\}$~yr, well within the run-to-run stochastic scatter. We verify the same insensitivity in our own runs using a subset with $t_{\rm disk} \in \{100,10000\}$~yr, presented in Appendix~\ref{app:t_disk_res}.

Jets are launched via a particle-spawning method \citep{2020torrey, 2021su,2024Su}, in which new gas resolution elements are created to represent the jet material at a height $r_{\rm inj}$ along the $z$-axis. The spawned particles are assigned an initial mass $m_{\rm j} = 0.1\,m_{\rm g,min}$, temperature $T_{\rm j} = 10^4\,\mathrm{K}$, and velocity $v_{\rm j}$, which together set the jet specific energy. The jet velocity is specified at the outset of each run and varies across our simulation suite (see Table~\ref{table:sims}). To ensure exact conservation of linear momentum, two particles are spawned simultaneously in opposite $\pm z$-directions whenever the accumulated jet mass flux reaches twice $m_{\rm j}$. A spawned particle is eligible to de-refine only once its speed has fallen below $0.1\,v_{\rm j}$, preserving the mass resolution of the jet beam throughout its propagation. 

The spawn height is set dynamically as $r_{\rm inj} \approx 0.125\,h_{\rm BH}$, where $h_{\rm BH} \sim (m_{\rm g,min}/\rho_0)^{1/3}$ is the local inter-particle spacing near the BH. As shown by \citet{2021pittard}, an energy-driven bubble forms only if the ram or thermal pressure at the edge of the injection region exceeds the ambient pressure, which sets a maximum injection radius $r_{\rm inj,max} = (\dot M_jv_j/4\pi P_0)^{1/2}$. For our fiducial run, with $P_0 = \rho_0c_s^2 \approx 2.4\times10^{-6}\,\mathrm{dyn\,cm^{-2}}$ and a regulated $\dot M_j \sim 10^{-4}\dot M_{\rm B}$, this gives $r_{\rm inj,max} \approx 2.1$\,mpc against $r_{\rm inj}\approx1.9\times10^{-2}$\,mpc. The criterion is therefore satisfied by two orders of magnitude, and, consistent with this, we find that the regulated state is unchanged across a factor of four variation in $m_{\rm g,min}$ (Appendix~\ref{app:t_disk_res}), over which $r_{\rm inj}$ varies by $60$ per cent.

Although set numerically, the jet velocity itself is a product of the unresolved launching region. For an outflow launched at some radius $r_l$ the natural velocity scale is the local escape velocity, $v_j\sim(2GM_{\rm BH}/r_{\rm l})^{1/2}$. For a $10\,M_\odot$ BH our fiducial $v_j = 10^{4}\,{\rm km\,s^{-1}}$ corresponds to $r_{\rm l}\simeq9\times10^{-7}$\,mpc, and the suite as a whole, $v_j = 10^{3}$--$3\times10^{4}\,{\rm km\,s^{-1}}$, spans $r_{\rm l}\simeq10^{-7}$--$9\times10^{-5}$\,mpc. Even the widest of these lies two orders of magnitude inside the spawn height $r_{\rm inj}\approx1.9\times10^{-2}$\,mpc and three orders of magnitude inside the sink radius $r_{\rm acc}\simeq0.13$\,mpc. The jet velocity therefore encodes physics far below our resolution and is properly treated as a free parameter of the sub-grid model, which we vary across the suite rather than deriving from the accretion flow.
 
The mass flux carried by the jet is set by a fixed feedback mass-loading factor,
\begin{equation}
    \dot M_j = \eta_{\rm m,fb}\,\dot M_{\rm BH},
\label{eq:eta_mfb}
\end{equation}
where $\dot M_{\rm BH}$ is the rate at which the $\alpha$-disk reservoir drains. We adopt $\eta_{\rm m,fb}\simeq1$ throughout (Table~\ref{table:sims}), so that the jet carries away as much mass as reaches the BH and roughly half of the captured gas is relaunched rather than accreted. We acknowledge that the most physical treatment would set the loading dynamically from the instantaneous feeding rate, following the trapping argument of \S\ref{subsec:super-Eddington_accretion}, in which $f_{\rm eject}$ falls as the cocoon drives the supply towards Eddington, but do not apply that approach here for simplicity and leave it for future simulations.

We do not expect the choice of $\eta_{\rm m,fb}$ to significantly change results. As \citet{2023su} show, it is the jet mass flux rather than the accretion rate that the momentum balance regulates. The simulations therefore constrain $\dot M_j$ directly, while $\eta_{\rm m,fb}$ rescales the simulated accretion rate. Taking the simulated outflow mass flux as the primary result, we can infer a physically motivated $\dot{M}_{\rm BH}$ through the trapping prescription of \S\ref{subsec:super-Eddington_accretion}, a procedure we set out in full in \S\ref{sec:discussion}.\footnote{Within a run, $\eta_{\rm m,fb}$ enters through the BH mass it accumulates, and that accumulation is negligible. At the fiducial suppression the BH grows at a rate $\dot{M}_{\rm BH, sim} = \dot{M}_j\sim10^{-4}\dot M_{\rm B}\simeq10^{-8}\, \rm M_\odot\,yr^{-1}$. Sustaining that rate for $10^{6}$\,yr would add only $\sim0.02\,M_\odot$. The impact of BH growth on the simulation results can thus be neglected and a different $\eta_{\rm m,fb}$ merely rescales the inferred accretion rate without altering the regulated outflow.}

\begin{table*}
\centering
\footnotesize
\setlength{\tabcolsep}{8pt}
\resizebox{\textwidth}{!}{%
\begin{tabular}{lccccccccc}
\toprule
Simulation & $L_{\rm box}$ & $\eta_{\rm m,fb}$ & $m_{\rm g,min}$ & $v_{\rm j}$ & $\rho_0$ & $T_0$ & $T_{\rm floor}$ & $Z$ & $t_{\rm acc}$ \\
 & ($10^{-3}$\,pc) & & ($M_\odot$) & (km\,s$^{-1}$) & (g \, cm$^{-3}$) & (K) & (K) & ($Z_\odot$) & (yr) \\
\midrule
\multicolumn{10}{l}{\textit{Fiducial}}\\
\cmidrule(l){1-10}
fid.no\_cool & $100$ & $0.96$ & $10^{-6}$ & $10^{4}$ & $2.0 \times 10^{-17}$ & $3000$ & $3000$ & No cooling & $10^{3}$ \\
fid.lo\_met  & $100$ & $0.96$ & $10^{-6}$ & $10^{4}$ & $2.0 \times 10^{-17}$ & $3000$ & $1741$ & $10^{-5}$ & $10^{3}$ \\
fid.hi\_met  & $100$ & $0.96$ & $10^{-6}$ & $10^{4}$ & $2.0 \times 10^{-17}$ & $3000$ & $1741$ & $1$ & $10^{3}$ \\
\midrule
\multicolumn{10}{l}{\textit{Varying jet velocity $v_{\rm j}$}}\\
\cmidrule(l){1-10}
vj1e8.lo\_met (hi\_met) & $100$ & $0.96$ & $10^{-6}$ & $10^{3}$ & $2.0 \times 10^{-17}$ & $3000$ & $1741$ & $10^{-5}$ ($1$) & $10^{3}$ \\
vj3e8.lo\_met (hi\_met) & $100$ & $0.96$ & $10^{-6}$ & $3\times10^{3}$ & $2.0 \times 10^{-17}$ & $3000$ & $1741$ & $10^{-5}$ ($1$) & $10^{3}$ \\
vj3e9.lo\_met (hi\_met) & $100$ & $0.96$ & $10^{-6}$ & $3\times10^{4}$ & $2.0 \times 10^{-17}$ & $3000$ & $1741$ & $10^{-5}$ ($1$) & $10^{3}$ \\
\midrule
\multicolumn{10}{l}{\textit{Varying ambient temperature $T_0$}}\\
\cmidrule(l){1-10}
T6e3.lo\_met (hi\_met)   & $100$ & $0.96$ & $10^{-6}$ & $3\times10^{4}$ & $2.0 \times 10^{-17}$ & $6000$  & $3482$  & $10^{-5}$ ($1$) & $10^{3}$ \\
T1.2e4.lo\_met (hi\_met) & $100$ & $0.96$ & $10^{-6}$ & $3\times10^{4}$ & $2.0 \times 10^{-17}$ & $12000$ & $6964$  & $10^{-5}$ ($1$) & $10^{3}$ \\
T2.4e4.lo\_met (hi\_met) & $100$ & $0.96$ & $10^{-6}$ & $3\times10^{4}$ & $2.0 \times 10^{-17}$ & $24000$ & $13928$ & $10^{-5}$ ($1$) & $10^{3}$ \\
\midrule
\multicolumn{10}{l}{\textit{Varying ambient density $\rho_0$}}\\
\cmidrule(l){1-10}
n3e7.lo\_met (hi\_met) & $100$ & $0.98$ & $10^{-6}$ & $10^{4}$ & $6.1 \times 10^{-17}$ & $3000$ & $1741$ & $10^{-5}$ ($1$) & $10^{3}$ \\
n1e8.lo\_met (hi\_met) & $100$ & $0.99$ & $10^{-6}$ & $10^{4}$ & $2.0 \times 10^{-16}$        & $3000$ & $1741$ & $10^{-5}$ ($1$) & $10^{3}$ \\
n3e8.lo\_met           & $80$ & $1.0$ & $10^{-6}$ & $10^{4}$ & $6.1 \times 10^{-16}$ & $3000$ & $1741$ & $10^{-5}$ & $10^{3}$ \\
\midrule
\multicolumn{10}{l}{\textit{Varying accretion-disk draining time $t_{\rm acc}$}}\\
\cmidrule(l){1-10}
tacc1e2.hi\_met          & $100$ & $0.96$ & $10^{-6}$ & $10^{4}$ & $2.0 \times 10^{-17}$ & $3000$ & $1741$ & $1$ & $10^{2}$ \\
tacc3e2.lo\_met (hi\_met) & $100$ & $0.96$ & $10^{-6}$ & $10^{4}$ & $2.0 \times 10^{-17}$ & $3000$ & $1741$ & $10^{-5}$ ($1$) & $3\times10^{2}$ \\
tacc3e3.lo\_met (hi\_met) & $100$ & $0.96$ & $10^{-6}$ & $10^{4}$ & $2.0 \times 10^{-17}$ & $3000$ & $1741$ & $10^{-5}$ ($1$) & $3\times10^{3}$ \\
tacc1e4.lo\_met          & $100$ & $0.96$ & $10^{-6}$ & $10^{4}$ & $2.0 \times 10^{-17}$ & $3000$ & $1741$ & $10^{-5}$ & $10^{4}$ \\
\midrule
\multicolumn{10}{l}{\textit{Varying gas mass resolution $m_{\rm g,min}$}}\\
\cmidrule(l){1-10}
mgmin5e-7.hi\_met & $100$ & $0.96$ & $5\times10^{-7}$ & $10^{4}$ & $2.0 \times 10^{-17}$ & $3000$ & $1741$ & $1$ & $10^{3}$ \\
mgmin2e-6.hi\_met & $100$ & $0.96$ & $2\times10^{-6}$ & $10^{4}$ & $2.0 \times 10^{-17}$ & $3000$ & $1741$ & $1$ & $10^{3}$ \\
\bottomrule
\end{tabular}%
}
\caption{Parameters of the simulation suite. Rows are grouped by the parameter being varied; within each group all other parameters are held at their fiducial values. Entries of the form ``lo\_met (hi\_met)'' denote a pair of runs identical except for metallicity, with $Z$ given as ``low (high).''}
\label{table:sims}
\end{table*}

% ----------------------------------------------------------------------------------------------------
% sec:methods
% subsec:setup_post_processing
% ----------------------------------------------------------------------------------------------------

\subsection{Post-processing}\label{subsec:setup_post_processing}

When post-processing our simulation data we use the following definitions to distinguish different cocoon phases. Jet particles are spawned with mass $m_{\rm jet} = 0.1\, m_{\rm g, min}$ and are identified in post-processing by the criterion $m_i \leq m_{\rm g, min}$. The remaining gas is partitioned by temperature into the hot interior ($T > T_{\rm hot}$, with $T_{\rm hot} = 10^6$ K), the warm phase ($T_{\rm warm} <  T \leq T_{\rm hot}$, with $T_{\rm warm} = 4\times10^4$ K), and the mixed phase ($T_{\rm mix} < T \leq T_{\rm warm}$, where $T_{\rm mix} = 1.2 ~ T_0$).  The two upper bounds are held fixed across the suite rather than scaled to the ambient conditions, so that the same absolute temperature ranges define the hot and warm phases in every run and the phases remain directly comparable between them. Only the lowest bound tracks the ambient state, since it exists to separate gas that has been heated from gas still at the floor. The cocoon is the union of these three heated phases. A shell phase captures cooler outflowing gas ($T\leq T_{\rm mix}$, $v_r \geq v_{\rm out}$, where $v_r$ is the radial velocity of the gas particle and $v_{\rm out} = c_s$ by default). The remaining gas is classified as ambient. 

For each phase we extract spatial and thermodynamic quantities. The maximum spatial extents are the cylindrical radius $R_{\rm max, cyl}$ and height $z_{\rm max}$, computed as volume-weighted, $\langle X \rangle = \sum_i X_i V_i / V_{\rm tot}$,  90th-percentile quantiles of the particle cylindrical radii ($R_{\rm cyl, i}$) and heights $|z_i|$, where $V_i = m_i/\rho_i$ is the effective volume of each resolution element and $\rho_i$ the kernel density stored in the snapshot.

The mean pressure and number density are also volume-weighted, which avoids biasing these quantities---particularly in the hot cocoon interior---toward the colder, denser particles at the phase boundary. We weight the average temperature by mass, $\langle T \rangle_m = \sum_i T_i m_i / \sum_i m_i$. Since the specific internal energy of each particle is $u_i = k_B T_i / (\gamma - 1)\mu m_p$, the thermal energy is $E_{\rm th} = \sum_i m_i u_i \propto \sum_i m_i T_i$, so the mass-weighted mean is the temperature the gas would have if the parcel were fully mixed at fixed total internal energy.

The cooling luminosity of each particle is $\dot{E}_i = n^2_{H, i} \Lambda_i V_i$, where $\Lambda_i$ is the cooling function stored directly in the snapshot and $n_{H,i}= \rho_i X_H/m_p$ where $X_H = 0.76$. The phase cooling luminosity and thermal energy are then $L_{\rm cool} = \sum_i \dot{E}_i$ and $E_{\rm th} = \sum_i m_i u_i$, giving a cooling time $t_{\rm cool} = E_{\rm th}/L_{\rm cool}$.

% ----------------------------------------------------------------------------------------------------
% sec:simulation_results
% ----------------------------------------------------------------------------------------------------

\section{Simulation Results}\label{sec:simulation_results}

We build our analysis around a fiducial set-up, with the ambient density and temperature set to $\rho_0 = 2 \times 10^{-17} ~\rm{g\,cm^{-3}}$ and $T_0 = 3\times10^3 ~\rm{K}$, respectively and jet velocity $v_j = 10^4 ~ \rm{km\,s^{-1}}$ (see Table~\ref{table:sims}). We examine this model under three cooling treatments: an adiabatic model in which the injected energy cannot be radiated (\S\ref{subsec:no_cooling_results}), and two radiative models that bracket the metallicity range $Z = 10^{-5} ~\rm{Z}_\odot$ and $Z = \rm{Z}_\odot$ (\S\ref{subsec:cooling_results}). These three treatments let us isolate the role of cooling. The no-cooling run forces adiabaticity by construction, while the two metallicities sample realistic cooling. In each case the regulated rate follows from the same momentum balance at $R_{\rm B}$ \citep{2023su} -- the cocoon's outward momentum flux against the inflowing Bondi momentum -- with cooling setting only what supplies the cocoon's side of that balance, thermal pressure retained in the interior or direct jet thrust. Having established this picture in the fiducial model, we study a broader suite, varying jet velocity, effective temperature, and ambient density (\S\ref{subsec:results_sim_suite}).

% ----------------------------------------------------------------------------------------------------
% sec:simulation_results
% subsec:no_cooling_results
% ----------------------------------------------------------------------------------------------------

\subsection{No-cooling run: a quasi-steady, mixing-regulated cocoon}
\label{subsec:no_cooling_results}

\begin{figure}
    \centering
\includegraphics[width=0.93\linewidth]{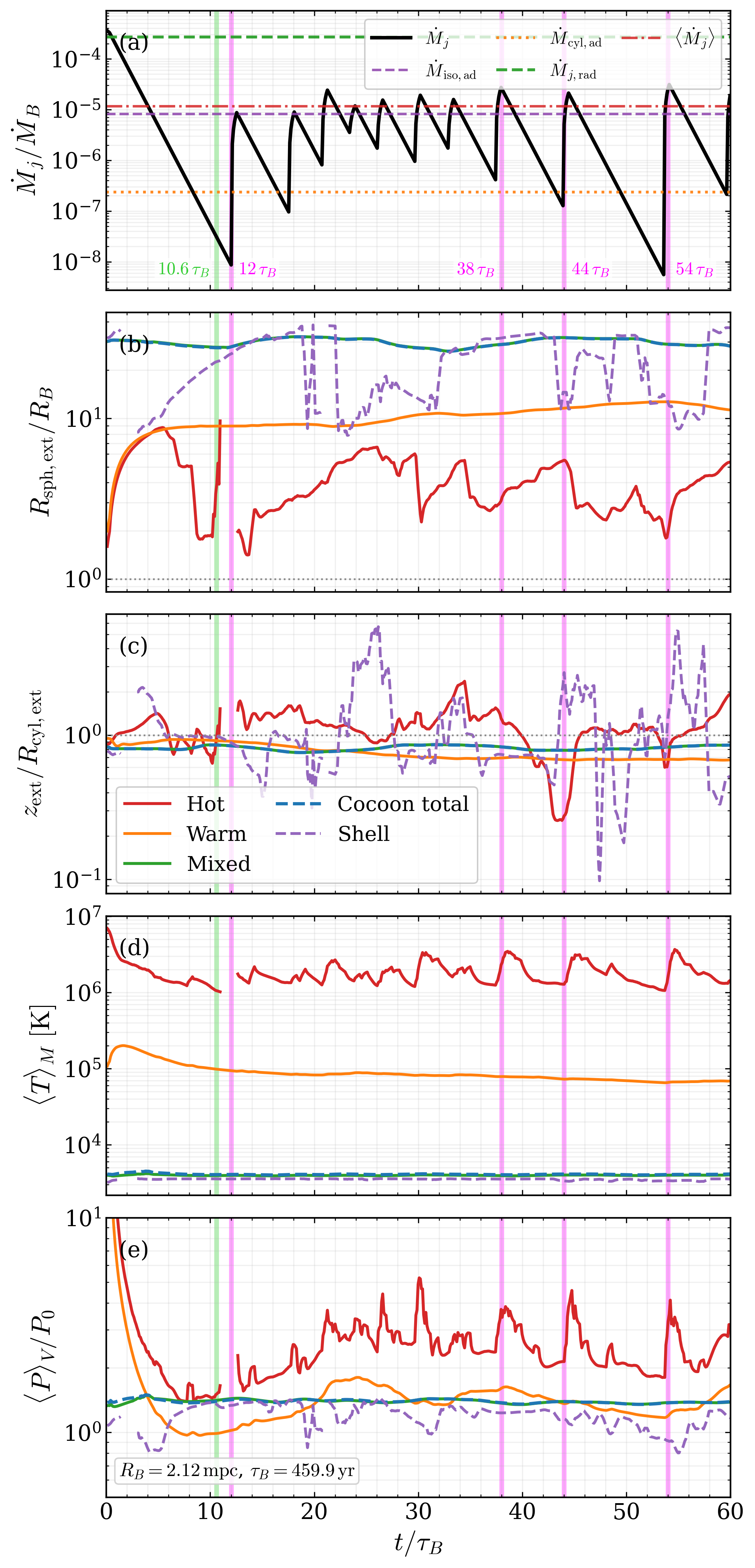}
    \caption{Time evolution of the fiducial no-cooling run over the first $60\,\tau_B\simeq 27.6\,\rm kyr$. Panel (a) shows the jet mass flux $\dot{M}_j$ normalized by the Bondi rate. Horizontal lines mark analytic estimates for the regulated mass flux, including the isotropic adiabatic estimate (purple, dashed), the momentum-regulated scaling [Eq.\ref{eq:mom_driven_regulated}] (green, dashed), and the cylindrical, adiabatic expansion estimate (orange, dotted). Panels (b)--(e) show the external radius, mass, mass-weighted temperature, and volume-weighted pressure of the cocoon phases: hot interior, warm phase, mixed/compressed gas, and the total cocoon. Vertical lines mark times discussed in the text. The green line at $10.6\,\tau_B$ is the snapshot shown in Fig.~\ref{fig:NO_COOL_snap_100}, while the pink lines at $12$, $38$, $44$ and $54\,\tau_B$ mark several, near-vertical rises in $\dot M_j$ (panel a).}
    \label{fig:NO_COOL_TS}
\end{figure}

\begin{figure}
    \centering
\includegraphics[width=0.95\linewidth]
{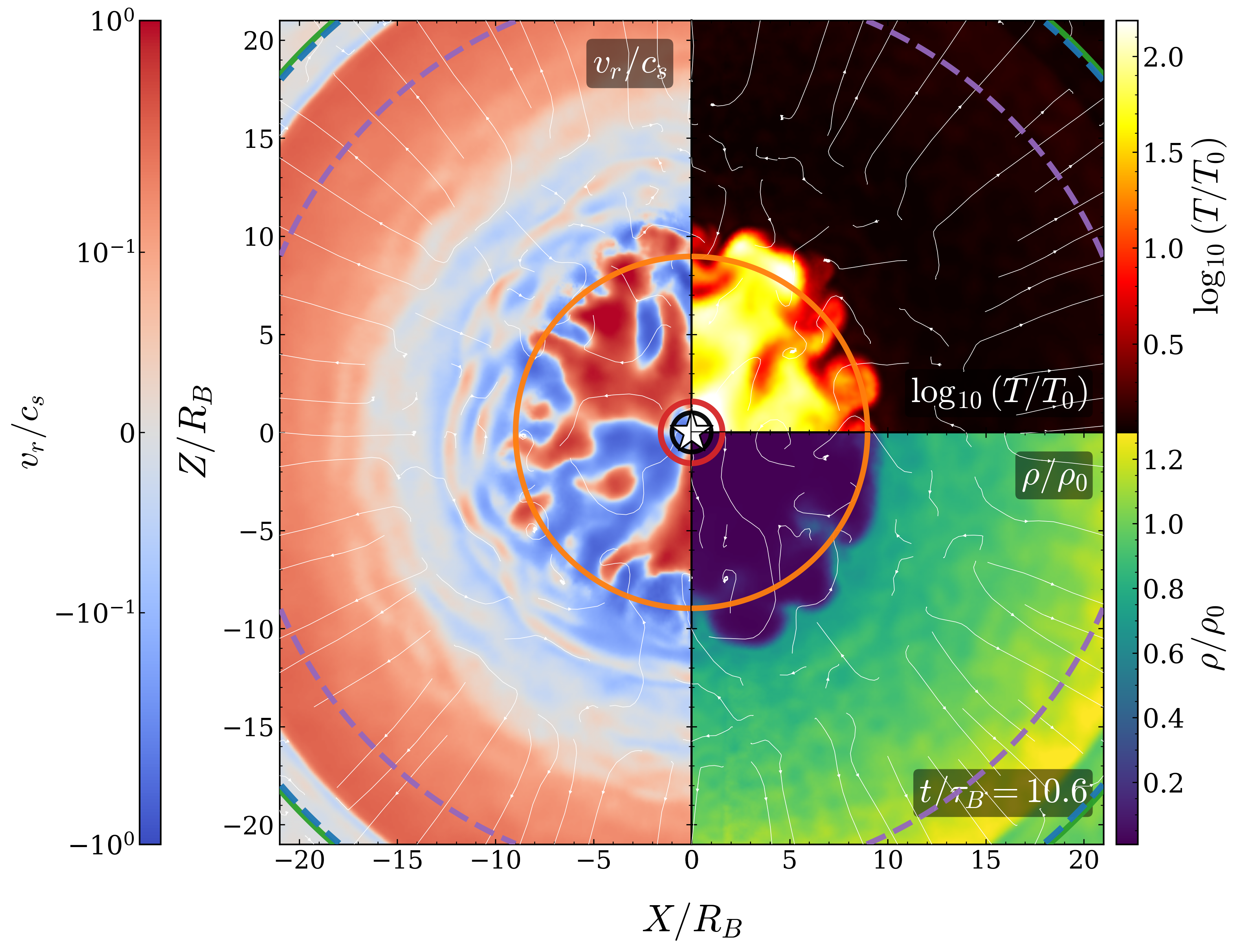}
 \caption{Snapshot of the fiducial no-cooling jet run at $t\simeq10.6\, \tau_B$ (green vertical line in Fig.~\ref{fig:NO_COOL_TS}) shown as an axisymmetric split-panel slice through the BH. The left half-plane shows the radial velocity normalized by the ambient sound speed $v_r/c_s$, while the upper-right and lower-right quadrants show $\log_{10}(T/T_0)$ and $\log_{10}(\rho/\rho_0)$, respectively. White streamlines indicate the projected gas flow in the slice and the white star marks the BH position. The black circle marks $R_B$, while the colored circles mark the radii of the hot interior, warm phase, mixed gas, and total cocoon -- as defined in \S\ref{subsec:setup_post_processing} -- and using the same colour convention as in Fig.~\ref{fig:NO_COOL_TS}.} 
    \label{fig:NO_COOL_snap_100}
\end{figure} 

We begin with the adiabatic run, in which radiative cooling is disabled. This run is closest to the Weaver-like bubble: with no radiative channel, the injected power can only be held in the interior as thermal pressure or exported mechanically. It is this pressure -- rather than direct momentum flux -- which is balanced against the inflowing Bondi momentum at $R_{\rm B}$. Disabling cooling thus isolates the regulation from any radiative channel. 

In Fig.~\ref{fig:NO_COOL_TS} (panels b-e) we show the radius, phase masses, temperature and pressure of all cocoon components plotted over $60\,\tau_B$, where $\tau_B=R_{\rm B}/c_s \simeq 460 \, \rm yr$ is the Bondi crossing time. These parameters settle to approximately constant values, fluctuating only at small amplitude around well-defined means. The jet mass flux -- equal to the simulated BH growth rate by construction -- likewise settles to a regulated value far below the Bondi rate, $\dot{M}_{\rm BH}\sim 10^{-5} ~\rm{\dot{M}_B}$ (panel a). The cocoon isotropizes early, and evolves as a spherical bubble. 

In the absence of cooling, the natural expectation is a Weaver-type adiabatic bubble \citep{1977weaver}, in which the cocoon expands self-similarly, $R \propto t^{3/5}$. But the saturation of our cocoon at a fixed size and pressure indicates that injected energy is neither stored as internal thermal energy nor spent on $P \mathrm{d}V$ work against the ambient medium, and must be released by some other mechanism. With radiative losses switched off by construction, the only channel available is the mechanical redistribution of energy by mixing at the cocoon interface, which advects hot-phase enthalpy into the cooler shell. 

This expectation is borne out by the instantaneous structure of the flow. Fig.~\ref{fig:NO_COOL_snap_100} shows characteristic Kelvin-Helmholtz (KH) spirals along the outer boundary of the warm phase (orange circle), where warm gas shears against the much denser and cooler mixed phase. The signed-velocity panel (left) resolves three zones across the boundary of the warm phase. Interior to it the radial velocity is stochastic with no coherent sign, the signature of a turbulent layer continuously processing hot gas into the shell. Exterior the warm phase boundary the velocity turns inward, as gas mass-loaded at the interface and denser than the interior at comparable pressure drains back towards the center. At $\sim 17 \, R_{\rm B}$ from the BH, there is a sharp transition to outward motion characterizing shell gas (purple, dashed circle), where mixing gives way to momentum transfer and the sweeping up of ambient material. 

The mechanics of such an interface have been characterized in detail for turbulent mixing layers \citep{1990begelman, 2020fielding} and radiative wind-blown bubbles \citep{2021lancaster_a, 2021lancaster_b, 2024lancaster}. There, KH mixing populates an intermediate-temperature layer at the contact discontinuity. This mixed gas radiates efficiently near the peak of the cooling curve, and the layer thickness self-adjusts until the hot-phase mixing time matches the cooling time and fixes the rate at which interior gas is processed. The interior pressure then follows from the balance between the injected power and the radiative losses at the interface. With cooling disabled, intermediate temperature gas in our simulation cannot radiate, so the mixed enthalpy is not dissipated but advected mechanically into the shell, where it is retained as thermal and bulk kinetic energy. The energy-balance framework is, however, unchanged, and in steady state the injected power must equal the enthalpy flux carried across the interface, 
\begin{equation}
    \dot{E}_{\rm in} = \frac{\gamma}{\gamma -1} P_h \, \langle v_{\rm out}\rangle \, A_{\rm int}.
\label{eq:nocool_energy_balance}
\end{equation}
Here $\langle v_{\rm out}\rangle$ is the mean entrainment velocity and $A_{\rm int}$ is the interface area. In place of the cooling curve setting the processing rate, the entrainment velocity is fixed kinematically by the turbulence, $\langle v_{\rm out} \rangle \simeq \mathcal{E} \, c_h$, with $c_h$ the hot-phase sound speed and $\mathcal{E}$ the dimensionless entrainment efficiency. 

The pressure to which the hot phase settles is fixed not by the interior energetics but by accretion feedback. The volume-weighted hot-phase pressure oscillates between $P_h \simeq 2\,-\,4\,  P_0$ (Fig.~\ref{fig:NO_COOL_TS}). The lower bound of this range is precisely the pressure predicted by the pressure-balance argument described in \citet{2023su}, that is, momentum balance gives $P_h = P_0 + \rho_0 \dot R^2$, so that $P_h \simeq 2P_0$ is identical to the condition $\dot R = c_s$ at $R_B$. At the Bondi radius this is the marginal state in which the cocoon edge sits at the sonic point of the accretion flow, neither overrun by infall nor advancing. Equivalently, $P_h \simeq 2P_0$ is the overpressure at which the ambient sound speed is approximately the local free-fall velocity, balancing the gravitational pull on the swept-up shell against the interior pressure at the gravitational-capture radius. 

The accretion rate and corresponding jet mass flux are thereby set by the cocoon: $P_h > 2 P_0$ suppresses accretion and starves the injection that sustains the cocoon, while $P_h < 2 P_0$ reignites it. This cycle is illustrated in Fig.~\ref{fig:NO_COOL_TS}, where the pink vertical lines mark four bursts in $\dot{M}_j$ (panel a), each accompanied by a jump in the volume-weighted hot-phase pressure (panel e, red). An accretion event and jet outburst lead to a spike in pressure, which prevents further accretion. The repeating sawtooth pattern in $\dot{M}_j$ is a signature of the sub-grid reservoir rather than of the cocoon. That is, each near-vertical rise marks a capture event delivering gas to the $\alpha$-disk and each subsequent decay is that reservoir draining exponentially on $t_{\rm disk} = 10^3$\,yr. Note that the hot phase briefly disappears from panels (b) - (e) at $t \simeq 12\,\tau_{B}$ (first pink line), towards the end of one such draining phase, when injection is cut for long enough that existing hot gas mixes into the warm phase faster than it is replenished.

The stability of this quasi-steady behavior follows from the temperature dependence of the loss term. The enthalpy flux carried across the interface scales as $L_{\rm mix} \propto P_h \langle v_{\rm out}\rangle A_{\rm int} \propto \rho_h T_h^{3/2}A_{\rm int}$, therefore the energy loss increases super-linearly with $T_h$. A positive fluctuation in the injected power raises $T_h$, which in turn enhances both the entrainment velocity and, through the accompanying expansion, the interface area, so the energy is exported faster than it accumulates and the excess is removed. The hot phase thus acts as its own thermostat, and whereas the cooling curve provides the restoring term in the radiative case, here that role is played by the temperature sensitivity of the mechanical mixing. This closes the feedback loop and accounts for the small-amplitude oscillations about the steady state in Fig.~\ref{fig:NO_COOL_TS} rather than runaway heating or unbounded expansion. 

Combining the pressure closure with the entrainment rate yields a prediction for the equilibrium radius. Substituting $P_h = 2 P_0$, $\langle v_{\rm out} \rangle = \mathcal{E}c_h$, and $A_{\rm int} = 4 \pi R^2$ into  Eq.~\eqref{eq:nocool_energy_balance} yields a single radius, 
\begin{equation}
    R_{\rm eq} \simeq \left( \frac{\dot{E}_{\rm in}}{20 \pi \, P_0 \mathcal{E}c_h}\right)^{1/2}\,.
\label{eq:nocool_R_eq}
\end{equation}
Note that for fixed $\dot{E}_{\rm in}$,$R_{\rm eq}$ is a stable attractor. A large cocoon ($R>R_{\rm eq}$) is under-pressured and contracts while a small cocoon $R < R_{\rm eq}$ is over-pressured and expands. The regulated injection follows from the adiabatic, isotropic mass-flux relation in Eq.~\eqref{eq:mdot_regulated}, 
\begin{equation}
    \dot{E}_{\rm in} = \frac{\epsilon \dot{M}_j v_j^2}{2} = \frac{154}{27 \, e^{3/2} }c_s^2\dot{M}_B.
\label{eq:Linj_Reg}
\end{equation}
Inserting Eq.~\eqref{eq:Linj_Reg} into Eq.~\eqref{eq:nocool_R_eq} and using $\dot{M}_B \propto \rho_0 c_s R_B^2$ together with $P_0 \propto \rho_0 c_s^2$, the ambient density cancels and the equilibrium radius scales with the ratio of internal and ambient sound speeds 
\begin{equation}
    \frac{R_{\rm eq}}{R_B} \propto \left(\frac{c_s}{\mathcal{E} c_h}\right)^{1/2}\,.
\label{eq:Req_scaling}
\end{equation}
From our simulation, $T_h \simeq 2\times10^6~\rm K$ and $c_h \simeq169 \, \rm{km/s}$. The equilibrium radius of the hot phase is $R_{\rm eq, h} \simeq 2 \, R_{\rm B}$, yielding an entrainment efficiency $\mathcal{E} \simeq 10^{-3}$. 

% ----------------------------------------------------------------------------------------------------
% sec:simulation_results
% subsec:cooling_results
% ----------------------------------------------------------------------------------------------------

\subsection{Radiative runs: mixing-limited cooling}
\label{subsec:cooling_results}
 
\begin{figure*}
    \centering
\includegraphics[width=0.98\linewidth]{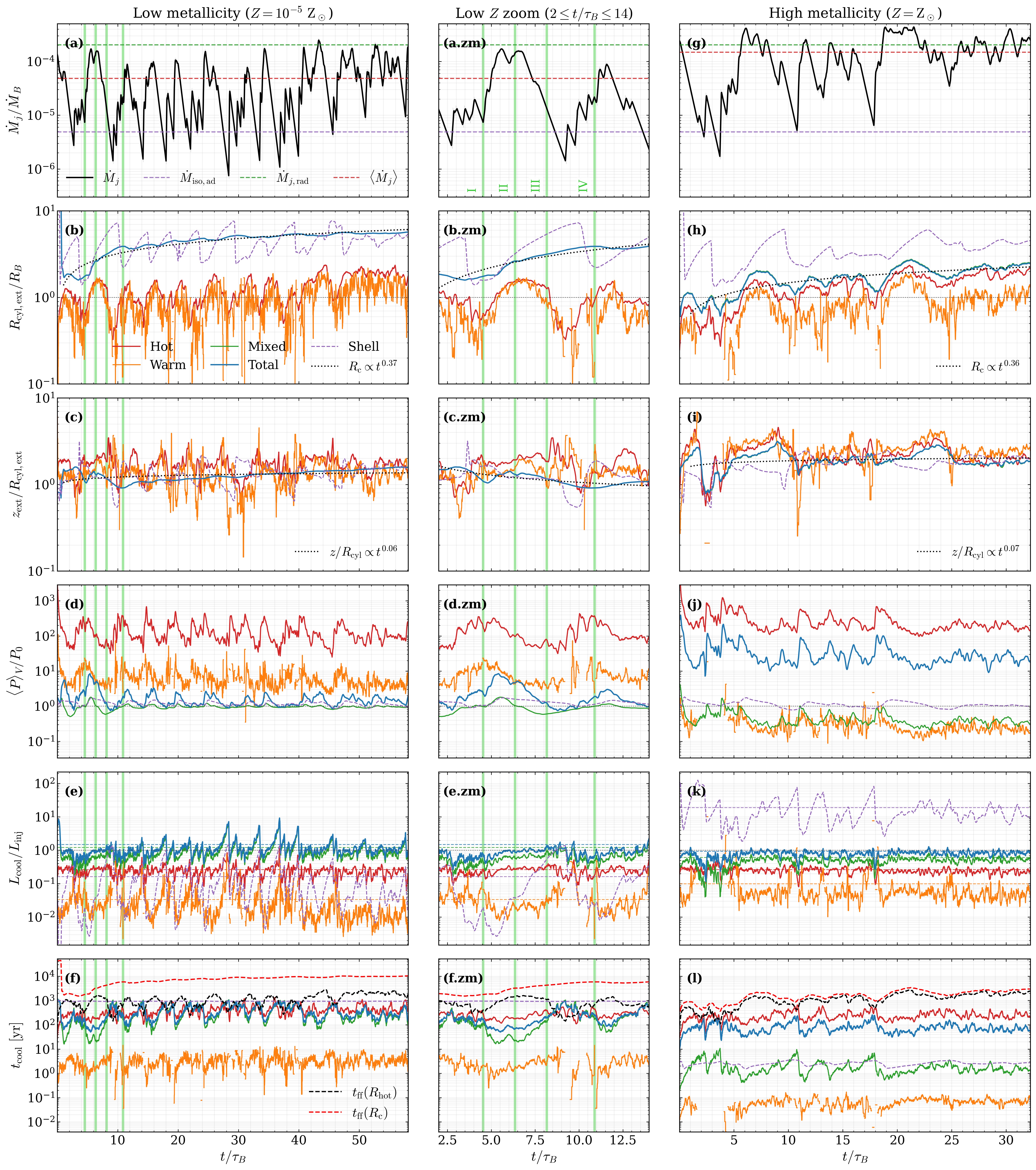}
 \caption{Time evolution of the low-metallicity ($Z = 10^{-5}\,Z_\odot$, left column, panels a--f), the same run over a narrower interval ($2 \le t/\tau_B \le 14$, middle column, panels a.zm--f.zm), and the solar-metallicity run ($Z = Z_\odot$, right column, panels g--l), with time in units of the Bondi crossing time $\tau_B$. (a, g) Jet mass flux $\dot{M}_j$ in units of the Bondi rate $\dot{M}_B$ (black solid), with its time average $\langle \dot{M}_j \rangle$ (red dashed), the adiabatic, isotropic momentum-balance prediction $\dot{M}_{\rm iso,ad}$ (purple dashed), and the momentum-driven rate $\dot{M}_{j,\rm rad}$ (green dashed), following the convention of the top panel of Fig.~\ref{fig:NO_COOL_TS}. (b, h) Cylindrical extent $R_{\rm cyl,ext}$ of each phase, normalized to $R_B$, with a power-law fit to the total-cocoon radius (black dotted). (c, i) Aspect ratio $z_{\rm ext}/R_{\rm cyl,ext}$, with power-law fit (black dotted). (d, j) Volume-averaged pressure $\langle P \rangle_V$ in units of the ambient pressure $P_0$. (e, k) Cooling luminosity $L_{\rm cool}$ normalized to the injected power $\dot{E}_{\rm in}$. (f, l) Cooling time $t_{\rm cool}$ of each phase, with the free-fall times of the hot interior $t_{\rm ff}(R_{\rm hot})$ (black dashed) and of the total cocoon $t_{\rm ff}(R_c)$ (red dashed). In panels (b)--(f) and (h)--(l), colored curves denote the hot interior (red), warm phase (orange), mixed phase (green), total cocoon (blue), and swept-up shell (purple dashed). Vertical green lines in the left and middle columns mark the four snapshots shown in Fig.~\ref{fig:low_z_snapshots}, labeled I--IV in the middle column, which together span a single oscillation cycle.}
    \label{fig:cool_ts}
\end{figure*}

\begin{figure}
    \centering
\includegraphics[width=0.99\linewidth]{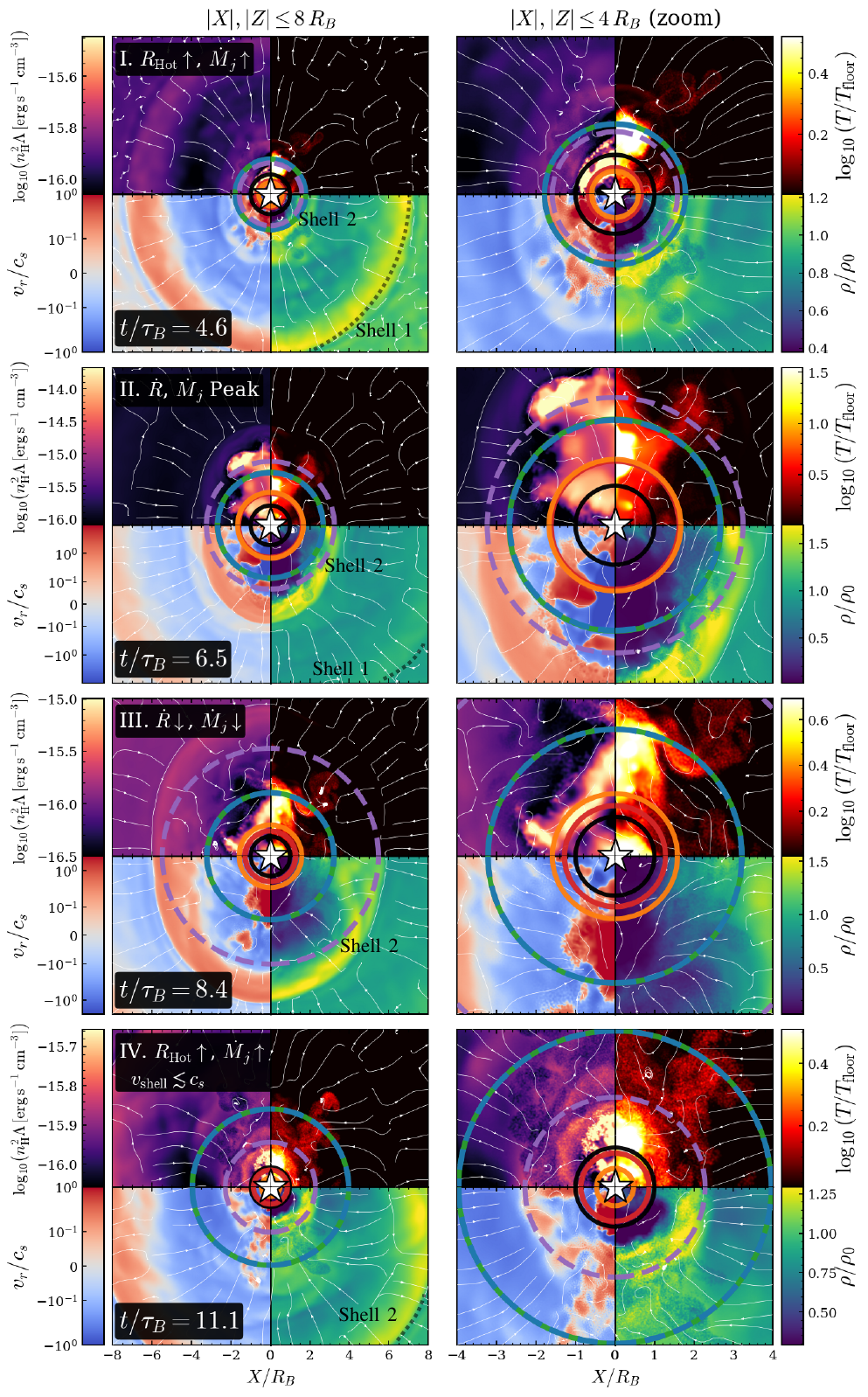}
 \caption{Four representative snapshots of the low-$Z$ simulation, shown from top to bottom at $t/\tau_B = 4.6$, 6.5, 8.4, and 11.1, where $\tau_B = 1040$ yr. Rows are labelled I--IV and follow one oscillation cycle, with the hot-phase radius and the jet mass flux rising together (I), peaking (II), declining (III), and rising again (IV). In row IV the shell velocity has fallen below the ambient sound speed, so the shell leaves the velocity-based classification of \S\ref{subsec:setup_post_processing} and is no longer followed. The same times are marked by vertical green lines in Fig.~\ref{fig:cool_ts}. The left column shows a thin slab containing the BH out to $|X|, |Z| \le 8\,R_B$ and the right column the same slab zoomed onto the jet and accretion region, $|X|, |Z| \le 4\,R_B$. The four quadrants show the cooling luminosity density, $\log_{10}(n_{\rm H}^2 \Lambda)$, in the upper-left; temperature, $\log_{10}(T/T_{\rm floor})$, in the upper-right; signed radial velocity, ${\rm sgn}(v_r)\log_{10}(|v_r|/c_s)$, in the lower-left; and density, $\rho/\rho_0$, in the lower-right. Here $T_{\rm floor}$, $\rho_0$, and $c_s$ are the ambient temperature, density, and sound speed. Each quadrant is scaled independently to emphasize the structure at that time. White streamlines trace the projected velocity field in the plotting plane, the white star marks the BH position, and the dashed white circle marks $R_B$. Colored dashed circles indicate the characteristic radii of the hot interior, warm phase, compressed gas, total cocoon, and swept-up shell. Labels in the left column identify the shell currently tracked, Shell 2, and the shell left by the previous outburst, Shell 1. A dotted gray line follows Shell 1 in the lower quadrants of rows I and II until it passes beyond the frame, and in row IV the same dotted gray line follows Shell 2, which has dropped below the velocity threshold and is no longer traced by the purple circle.}
    \label{fig:low_z_snapshots}
\end{figure}

We restore radiative cooling in runs fid.lo\_met and fid.hi\_met, distinguished by their sub-solar ($Z = 10^{-5}\,\rm Z_\odot$) and solar metallicities respectively. The full evolution of both runs is shown in Fig.~\ref{fig:cool_ts}, whose left and right columns follow the low- and solar-metallicity runs respectively, while the middle column zooms in on the interval $2 \le t/\tau_B \le 14$ of the fid.lo\_met run, its panels carrying an appended `.zm' on their labels. The panels track the jet mass flux (a, g), the radial extent and aspect ratio of each cocoon phase (b, c and h, i), the phase pressures (d, j), the cooling luminosity (e, k), and the characteristic cooling and free-fall times (f, l).

From panels (e) and (k) it is clear that the cooling luminosity of the cocoon total (blue) tracks the injected energy flux closely, with time averages of $\langle L_{\rm cool}/\dot{E}_{\rm in}\rangle = 0.99$ and $0.74$ in the low- and solar-metallicity runs (horizontal dashed lines). The cocoon thus radiates away the majority of the energy it is fed, shedding it nearly as fast as it arrives at the turbulent contact surface between the hot interior and the surrounding shell, where mixing exposes hot gas to cool, dense material with a very short cooling time. It may seem surprising that the
low-metallicity run radiates the larger fraction of injected energy between the two, since metal cooling would naively be expected to radiate more. The excess is, however, episodic rather than sustained, arriving in brief excursions above unity in panel (e) that coincide with cyclical bursts in $\dot{M}_j$, and does not indicate more efficient radiation in low-metallicity gas
generally.

The efficient radiation evident in both runs does not imply that the cocoon has relaxed into a fully radiative structure. In a fully radiative cocoon, energy conservation requires $\dot{E}_{\rm in}\simeq L_{\rm mix}\simeq L_{\rm cool}$ -- energy injected into the interior, carried through the contact discontinuity by mixing, and radiated away -- with the mixing layer self-adjusting until the two proceed in step, $t_{\rm mix}\sim t_{\rm cool}$. Our runs do not reach this balance. The warm, intermediate-temperature gas radiates almost on contact ($t_{\rm cool}\simeq4$ and $0.07$\,yr in the low- and high-metallicity runs, panels f, l), and although the hot interior cools more slowly ($t_{\rm cool}\simeq430$ and $240$\,yr respectively), estimating the mixing time as $t_{\rm mix}\simeq R_h/(\mathcal{E}c_h)$ with $R_h\simeq R_{\rm B}$ and $\mathcal{E}\simeq10^{-3}$ from the no-cooling run gives $t_{\rm mix}\simeq4.6\times10^{3}$\,yr (low-$Z$) and $3.1\times10^{3}$\,yr (solar), roughly an order of magnitude longer than $t_{\rm cool}$. This suggests that cooling sets the pace in the outer mixed region, which radiates as fast as it is fed, while the interior loses energy only as quickly as mixing delivers it to the interface and so retains part of the injected power rather than radiating it -- though we caution that the mixing estimate rests on an entrainment efficiency calibrated from the adiabatic run. What is robust is the resulting structure: a radiatively efficient surface bounding an interior that remains strongly over-pressured throughout (panels d, j). In holding on to part of its energy in this way the cocoon more closely resembles a partially radiative bubble \citep{1992koo_a, 1992koo_b, 2008matzner} -- intermediate between the energy-conserving (adiabatic) and fully radiative, momentum-driven limits -- than either extreme. Consistent with this, the time-averaged mass fluxes fall between the adiabatic momentum-balance criterion of \citet{2023su} ~[Eq.~\eqref{eq:mdot_regulated}] and its momentum-driven counterpart~[Eq.~\eqref{eq:mom_driven_regulated}] (panels a,g).

The cyclical bursts in the low-metallicity run are characterized by large-amplitude oscillations in both the jet mass flux (panel a) and the hot-interior radius (panel b), with bursts recurring on a period of $\sim5\times10^{3}$\,yr, or $\sim5\,\tau_{\rm B}$ with the Bondi crossing time of the radiative runs, $\tau_{\rm B}=1040$\,yr. Fig.~\ref{fig:low_z_snapshots} shows four snapshots of the low-metallicity run spanning one such cycle, each a thin slice of width $0.5\,R_{\rm B}$, centered on the BH. The left column shows the slices across $|X|,|Z| < 8\, R_B$ and the right column zooms in on $|X|,|Z| < 4\, R_B$. The time corresponding to each slice is marked by green vertical lines in the left and middle columns of the timeseries in Fig.~\ref{fig:cool_ts}.

The onset of each burst coincides with the free-fall time of the hot region, $t_{\rm ff}$, falling below its cooling time (dashed black line, Fig.~\ref{fig:cool_ts} panel (f)). The hot phase then expands rapidly and the jet mass flux rises with it (I), the two peaking together (II) before both decline (III). The expansion drives a density wave outward through the mixed phase and into the ambient medium, labeled Shell 2 in Fig.~\ref{fig:low_z_snapshots} to distinguish it from Shell 1, the shell left by the previous outburst, which is already coasting outward at the start of the cycle and passes beyond the frame between snapshots II and III. Shell 2 appears in Fig.~\ref{fig:cool_ts} panel (b) in the shell phase (purple, dashed curve) -- defined by gas with a radial velocity exceeding the ambient sound speed and $T<1.2\,T_{\rm floor}$ -- which we follow until the wave decelerates to the sound speed. At that point the gas falls outside the shell-phase velocity criterion and the purple circle jumps inward in Fig.~\ref{fig:low_z_snapshots} slice $IV$ . Note that the purple line indicates the shell radius defined in \S\ref{subsec:setup_post_processing} as a volume-weighted quantile over whatever gas currently satisfies the temperature and velocity criterion. When the leading front slows below $c_s$ it drops out of the sample and the shell boundary falls back onto the faster, more recently launched gas. The discontinuity thus marks the moment the outermost shell material leaves the diagnostic rather than any physical collapse. By snapshot IV the hot radius and jet mass flux have begun to rise again while Shell 2, still travelling outward and detached from the cocoon, has dropped below $c_s$ and is no longer traced by the purple curve, marked instead by the dotted grey line as Shell 1 was in snapshots I and II.

The detachment of the shell also accounts for the excursions of $L_{\rm cool}$ above $\dot{E}_{\rm in}$ noted above. Once Shell 2 pulls away from the cocoon the interior is no longer bounded by a dense, coherent shell, and the cocoon sheds stored thermal energy through a brief episode of interior mixing rather than through the steady interfacial layer that operates for the rest of the cycle. Radiating faster than it is fed, the cocoon depressurises, and the pressure drop is visible across the phases in panel (d), though not uniformly. The hot and warm phases stay well above $2\,P_0$ throughout, while the total cocoon, mixed phase and shell fall below it, so the loss is concentrated in the cooler, outer material rather than in the interior that sets the momentum balance. With the cocoon no longer able to hold the ram pressure of the inflow at $R_{\rm B}$, a capture episode follows and the jet mass flux spikes, restoring the pressure across all phases. The sequence closely parallels the accretion-driven pressure cycle of the adiabatic run (\S\ref{subsec:no_cooling_results}), with radiative losses at shell detachment playing the role that mechanical export through the interface plays there. 

This limit cycle is also a close, mechanical analogue of the radiation-regulated accretion duty cycle first identified in the radiation-hydrodynamic simulations of \citet{2009milosavljevic_b} and characterized in detail by \citet{2011park, 2012park}, in which photoionization feedback inflates an over-pressured H\,\textsc{ii} region that suppresses accretion until the region recombines and the surrounding gas refills. The parallels are suggestive -- a feedback-inflated, over-pressured bubble ultimately chokes the infall that drives it, so each burst is self-limiting and turns over near a pressure maximum. Once the supply is cut, the bubble can no longer sustain its over-pressure. In their case this leads to recombination and gravitational collapse that refills the mass-flux reservoir on a dynamical timescale, whereas here the shell coasts outward and detaches. Refill instead appears to proceed via infall at the shell's inner edge, where a region of bulk inward velocity becomes apparent by the end of the burst (Snapshot IV, Fig.~\ref{fig:low_z_snapshots}). This points to a somewhat different mechanism behind our cycle, even as the broad phenomenology is shared.

In the high-metallicity run the oscillatory behavior is far less pronounced, amounting to a perturbation on top of the regular expansion -- whose scaling matches the average expansion rate in the low-metallicity run ($R\propto t^{0.37}$). The mixed phase here has a much shorter cooling time and tracks the interior dynamically rather than evolving as a distinct reservoir. Compression waves still appear -- visible in the shell phase -- but a clear acceleration/burst cycle is seen only three times over $\sim 20\,\tau_B$, after which the oscillations largely damp away, and the accretion rate onto the BH correspondingly oscillates less.

The contrast between the two metallicities maps onto the two oscillation modes identified by \citet{2012park}. In their strong, low-duty-cycle Mode-I, bursts are driven by the collapse of neutral gas once the feedback-supported bubble is depleted, whereas in the mild, high-duty-cycle Mode-II the accretion rate is modulated by density waves launched from the bubble edge. Our low-metallicity run resembles the former -- large-amplitude bursts terminated by shell detachment and refilled by infall at the shell's inner edge -- while the solar metallicity run, in which a clean cycle appears only a few times before damping into residual compression waves, resembles the latter. We caution that the analogy is qualitative rather than quantitative: they move between modes along the product $M_{\rm BH}n_0$, whereas our runs share a common $M_{\rm BH}$ and $n_0$ and differ only in cooling efficiency through metallicity. In both cases, however, enhanced cooling damps the oscillation and raises the mean accretion rate, suggesting that it is the cooling efficiency, rather than the mechanics of the cocoon, that selects the mode. The no-cooling run, lacking any radiative channel, has no counterpart among these modes and instead settles onto the quasi-steady, mixing-regulated state of \S\ref{subsec:no_cooling_results}. We defer a mechanistic account of the transition to future work. The offset between our two runs is in any case modest -- both have $\langle \dot{M}_j/\dot{M}_{\rm B} \rangle \sim 10^{-4}$ and lie within the band bracketed by the adiabatic and momentum-regulated limits -- whereas the \citet{2012park} modes correspond to qualitatively distinct regulated states, Mode-I (strong, bursty oscillation) holding the mean at $\sim 1$ per cent of the Bondi rate and the other (Mode-II, mild oscillation) pinned near $\dot{\rm M}_{\rm Edd}$.

Finally, both runs settle to a stable, modestly elongated geometry. The cocoon radius grows as $R\propto t^{0.37}$ in both metallicities (Fig.~\ref{fig:cool_ts}, panels b, h), shallower than either the energy-conserving $R\propto t^{2/3}$ or the momentum-driven $R\propto t^{1/2}$, and the aspect ratio $z/R$ holds nearly constant throughout (panels c, i). Neither limiting expectation captures this scaling. An adiabatic cocoon at the high densities considered here should isotropise well before reaching the Bondi radius, while a fully radiative, momentum-driven cocoon, with its interior pressure radiated away, should likewise round out (\S\ref{subsec:momentum_driven}). Both limits therefore predict a sphere, yet our cocoons retain a persistent, if modest, elongation.
 
% ----------------------------------------------------------------------------------------------------
% sec:simulation_results
% subsec:results_sim_suite
% ----------------------------------------------------------------------------------------------------

\subsection{Parameter Space Exploration}\label{subsec:results_sim_suite}

\begin{figure*}
    \centering
\includegraphics[width=0.8\linewidth]{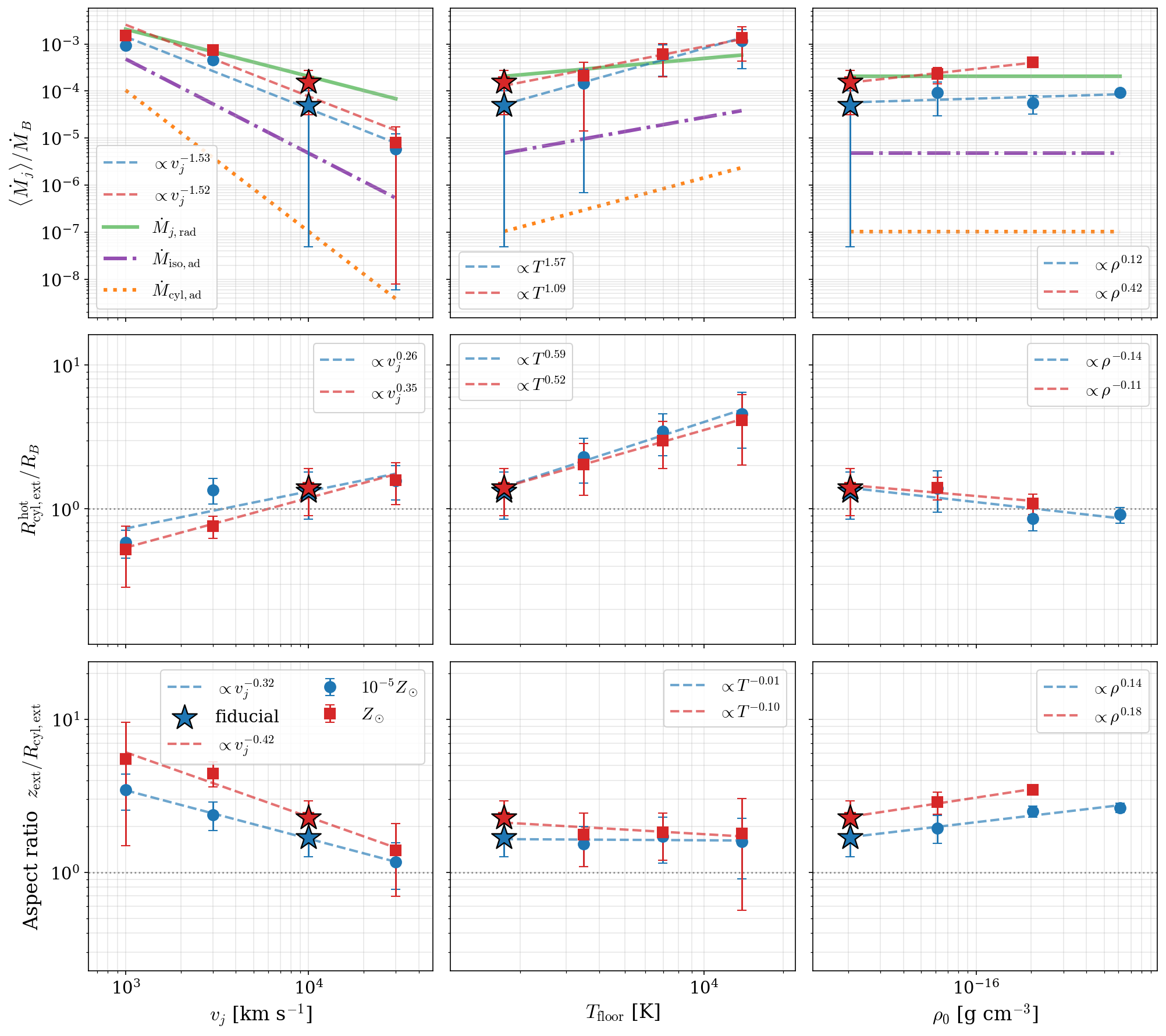}
 \caption{Summary of the parameter suite varying the jet velocity, ambient temperature, and ambient density. Columns show runs in which only one parameter -- $v_j$ (left), $T_{\rm floor}$ (middle), or $n_0$ (right) -- is varied relative to the fiducial model. Rows show the time-averaged accretion rate onto the embedded BH, $\langle \dot{M}_{j}\rangle /\dot{\rm M}_{\rm B}$ (top), the radial extent of the hot phase, $R_{\rm cyl, ext}^{\rm hot}/ \rm{R_B}$ (middle), and the associated aspect ratio $z/R_{\rm cyl, ext}^{\rm hot}$ (bottom). Blue circles denote low-metallicity runs, $Z = 10^{-5}~ \rm{Z}_{\odot}$, while red squares denote solar-metallicity runs, stars denote fiducial runs fid.lo\_met (blue) and fid.hi\_met (red). Points show time averages over the entire simulation and  errors bars indicate the minimum and maximum values seen. Dashed lines show power-law fits to each metallicity subset, with the fitted scalings labeled in each panel. In the accretion-rate row we also show analytic reference scalings: the momentum-regulated expectation (green), the isotropic adiabatic estimate $\dot M_{\rm iso, ad}$ (purple dashed), and the cylindrical adiabatic estimate $\dot M_{\rm cyl, ad}$ (orange dashed).
 }
    \label{fig:suite_grid}
\end{figure*}

We now turn to the time-averaged results of the full simulation suite. As in the fiducial runs, the cocoons surveyed here sit between the classically adiabatic and radiative regimes. These results are summarized in Fig.~\ref{fig:suite_grid}, whose columns each vary a single parameter about the fiducial set-up -- the jet velocity $v_j$ (left), the floor temperature $T_{\rm floor}$ (middle), and the ambient density $n_0$ (right) -- for both the low (blue) and high (red) metallicity cases.

The intermediate character of the regulation is clearest in the top row, which shows $\dot M_{ j}/\dot M_B$. Every point lies at or below the momentum-regulated expectation $0.6 (c_s/v_j)\dot M_B$ (green) and above the isotropic and cylindrical adiabatic estimates $\dot M_{\rm iso,ad}$ and $\dot M_{\rm cyl,ad}$ (purple and orange). The same is true of the jet-velocity scaling: the accretion rate follows $\langle\dot M_{ j}\rangle\propto v_j^{-1.5}$ at both metallicities, midway between the momentum-regulated $v_j^{-1}$ and the adiabatic $v_j^{-2}$. 

The remaining $\dot{M}_{\rm j}$ scalings are less clean. The density dependence is the most weakly constrained: the low-metallicity runs are essentially flat ($\propto n_0^{0.12}$), as expected if the regulated suppression depends only on $c_s/v_j$, whereas the solar-metallicity runs rise appreciably ($\propto n_0^{0.42}$). With only three densities sampled, we treat this difference as tentative, pending a finer grid. The temperature dependence is steeper than the momentum-regulated $T_{\rm floor}^{1/2}$ at both metallicities -- $\propto T_{\rm floor}^{1.57}$ (low-$Z$) and $T_{\rm floor}^{1.09}$ (solar), the former steeper even than the adiabatic linear scaling. Moreover, with four temperatures sampled per metallicity, this scaling is more robust than the density trend. The runs climb toward the momentum-regulated line at the highest $T_{\rm floor}$, but remain bracketed, within error, by the two analytic limits. It is the slope, not the normalization, that the simple regulation scalings fail to reproduce. 

This distinction matters when the fits are extrapolated. Within the suite the momentum-regulated line bounds every point from above, but it does so with shallower dependences on both temperature and density than the simulations exhibit. Over the much wider range of disk conditions shown in \S\ref{sec:AGN_params} the two cross: at the higher temperatures and densities reached at small $a$, the fitted jet mass fluxes rise above $\dot M_{j,\rm rad}$, even though they lie below it everywhere in the sampled parameter range. The jet velocity is the exception, being the one parameter whose measured slope is steeper than the analytic one. We discuss these scalings further in \S\ref{sec:discussion}.

\begin{figure*}
    \centering
\includegraphics[width=0.98\linewidth]{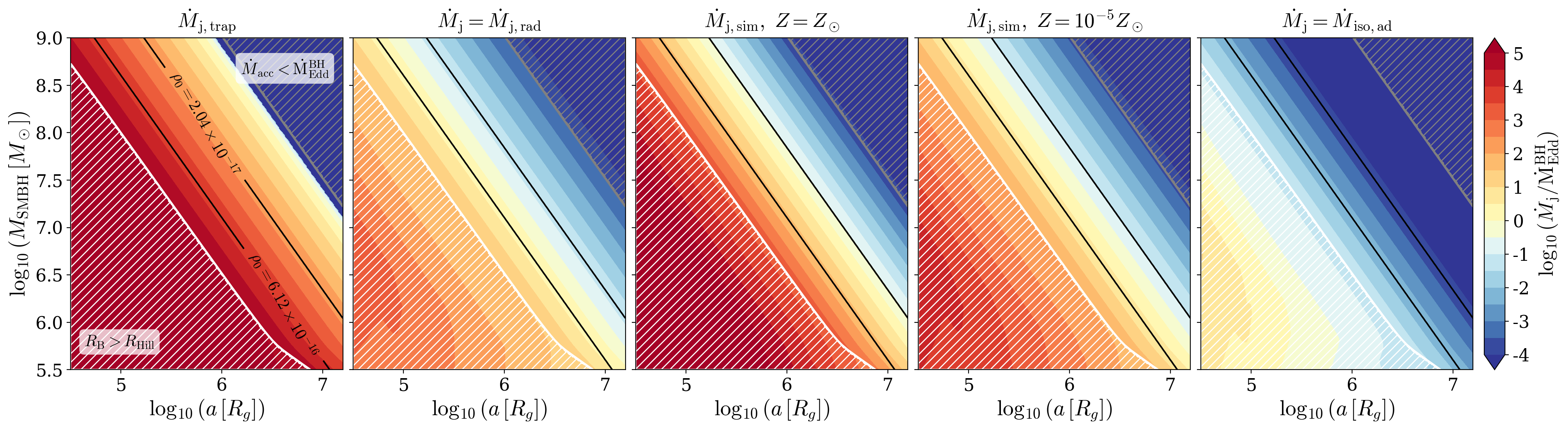}
 \caption{Regulated $\dot{M}_{\rm j}$ in units of $\dot{\rm M}_{\rm Edd}^{\rm BH}$, across the \citetalias{2003sirko} disk plane of central mass $M_{\rm SMBH}$ and orbital distance $a$, for a disk mass flux $\dot{M}_{\rm AGN} = \dot{\rm M}_{\rm Edd}^{\rm SMBH}$, viscosity $\alpha=0.1$, and assuming a fiducial jet velocity $v_{\rm j}=10^{4}\,{\rm km\,s^{-1}}$. Each panel applies a different prescription for $\dot{M}_j$, from left to right: the trapping-only prediction, in which no cocoon forms and the jet relaunches the full fraction $f_{\rm eject}$ of the supply; the radiative moment-balance $\dot{M}_{\rm j,rad}$ (Eq.~\ref{eq:mom_driven_regulated});  the fitted simulation scalings at solar and at low metallicity (\S\ref{subsec:results_sim_suite}); and the isotropic, adiabatic limit (Eq.~\ref{eq:mdot_regulated}). White hatching marks the region $R_{\rm B} > R_{\rm Hill}$. Gray hatching marks where $\dot{M}_{\rm acc} < \dot{M}_{\rm Edd}^{\rm BH}$, and the radiation trapping that powers the outflow shuts off. Solid black contours trace the mid-plane densities sampled by our simulation suite. }
    \label{fig:mdotjet_5col_maps}
\end{figure*}

% ----------------------------------------------------------------------------------------------------
% sec:simulation_results
% subsec:results_sim_suite
% subsubsec:cocoon_size_shape
% ----------------------------------------------------------------------------------------------------

\subsubsection{Cocoon size and shape}
\label{subsubsec:cocoon_size_shape}

Two geometric features of the suite stand out, and neither matches free self-similar expansion. The first is that the cocoon radius tracks $R_{\rm B}$ across the full parameter range. $R_{\rm cyl,ext}/R_B$ stays of order unity in every run (Fig.~\ref{fig:suite_grid}, middle row), varying only weakly with jet velocity ($\propto v_j^{0.26\text{--}0.35}$), ambient temperature ($\propto T_{\rm floor}^{0.39\text{--}0.52}$), and ambient density ($\propto n_0^{-0.11\text{--}-0.14}$). Self-similar cocoon expansion has no preferred scale, so the pinning of $R_{\rm cyl,ext}$ at $\simeq R_B$ is itself the signature of regulation. 

The second is that the aspect ratio is nearly flat in every sweep, which we can test directly against the free-expansion prediction. The elongated self-similar law is the only free-expansion model with a non-trivial aspect ratio. Forming $z/R$ from the elongated scalings [Eqs.~\eqref{eq:theory_R_aniso}--\eqref{eq:theory_z_aniso}] gives $z/R\propto\dot M_j^{1/6}v_j^{-1/2}\rho_0^{-1/6}t^{-1/3}$, and eliminating the time in favor of the evaluation radius through $R(t)$ [Eq.~\eqref{eq:theory_R_aniso}] yields $z/R\propto\dot M_j^{1/4}v_j^{-1/4}\rho_0^{-1/4}R^{-1/2}$. Writing the regulated flux as $\dot M_j=(\dot M_j/\dot M_B)\dot M_B$ with $\dot M_B\propto\rho_0 c_s^{-3}$, and evaluating at $R\simeq R_B$ with $R_B\propto c_s^{-2}$, the ambient density cancels and the aspect ratio reduces to
\begin{equation}
    \frac{z}{R}\propto\left(\frac{\dot M_j}{\dot M_B}\right)^{1/4} c_s^{1/4}\,v_j^{-1/4}\left(\frac{R}{R_B}\right)^{-1/2},
\label{eq:zR_master}
\end{equation}
where $\dot M_j/\dot M_B$ is read from the top row of Fig.~\ref{fig:suite_grid}. Note that the ambient density has dropped out, so $z/R$ depends on $n_0$ only through the weak scaling with $R/R_{\rm B}$. The temperature dependence is nearly canceled as well, with $c_s^{1/4}\propto T_{\rm floor}^{1/8}$. 

Defining the per-variable logarithmic slopes of the three quantities fitted in Fig.~\ref{fig:suite_grid} (top, middle, and bottom rows),
\begin{equation}
s_x\equiv\frac{\mathrm{d}\ln(\dot M_j/\dot M_B)}{\mathrm{d}\ln x},\qquad
q_x\equiv\frac{\mathrm{d}\ln(R/R_B)}{\mathrm{d}\ln x},\qquad
\zeta_x\equiv\frac{\mathrm{d}\ln(z/R)}{\mathrm{d}\ln x},
\label{eq:slope_defs}
\end{equation}
and differentiating Eq.~\ref{eq:zR_master} with
$\mathrm{d}\ln c_s/\mathrm{d}\ln T_{\rm floor}=1/2$,
the predicted aspect-ratio slopes are
\begin{equation}
\hat\zeta_T=\frac{1}{4}s_T+\frac{1}{8}-\frac{1}{2}q_T,\qquad
\hat\zeta_n=\frac{1}{4}s_n-\frac{1}{2}q_n,\qquad
\hat\zeta_v=\frac{1}{4}s_v-\frac{1}{4}-\frac{1}{2}q_v.
\label{eq:zeta_pred}
\end{equation}
For temperature, the momentum, adiabatic, and measured accretion suppression branches $s_T=1/2$, $1$, and the fitted $1.57/1.09$ (low-$Z$/solar) combine with $q_T=0.39/0.52$ to predict $\hat\zeta_T$ between $+0.06$ and $+0.32$ (low-$Z$) and $-0.01$ and $+0.14$ (solar), against the measured $\zeta_T=-0.01$ and $-0.10$. For density, the analytic branches share $s_n=0$, so the prediction comes from the radius slope alone, $\hat\zeta_n=-\tfrac{1}{2}q_n\simeq+0.06$--$0.07$; folding in the measured $s_n=0.12/0.42$ raises it to $+0.10$ and $+0.16$, against the measured $+0.14$ and $+0.18$, so the weak observed rise comes mostly from the radius scaling. 

The jet velocity is where the prediction is most markedly different from measured scalings. Every branch predicts a steep slope, $\hat\zeta_v$ from $-0.63$ (momentum) through $-0.76$ (measured) to $-0.88$ (adiabatic) at low $Z$, and $-0.68$ to $-0.93$ at solar, yet the measured slopes are only $\zeta_v=-0.32$ and $-0.42$, roughly half as steep. Together with the pinning of $R_{\rm cyl,ext}$ at $R_B$, this result is in line with the shallow expansion noted in \S\ref{subsec:cooling_results}.

% ----------------------------------------------------------------------------------------------------
% sec:discussion
% ----------------------------------------------------------------------------------------------------

\section{Discussion}\label{sec:discussion}

The scaling relations assembled above -- the regulated accretion rate and its dependence on jet velocity, temperature, and density -- let us estimate how an embedded BH would grow, and whether its cocoon could break out of the disk, across the broader range of conditions in the outer disk. We pursue these estimates in the following section. We caution, however, that these estimates take the fitted relations at face value. While the simulations establish that the regulated rate sits between the adiabatic and momentum-driven limits and scales roughly as $v_j^{-1.5}$, the mechanism that sets this intermediate slope -- and the cocoon's departure from free self-similar expansion -- is not yet fully understood. We therefore present what follows as a preliminary mapping of the consequences, and emphasize that firmer predictions will require both a physical account of the regulation and wider parameter coverage.

\begin{figure*}
    \centering
\includegraphics[width=0.98\linewidth]{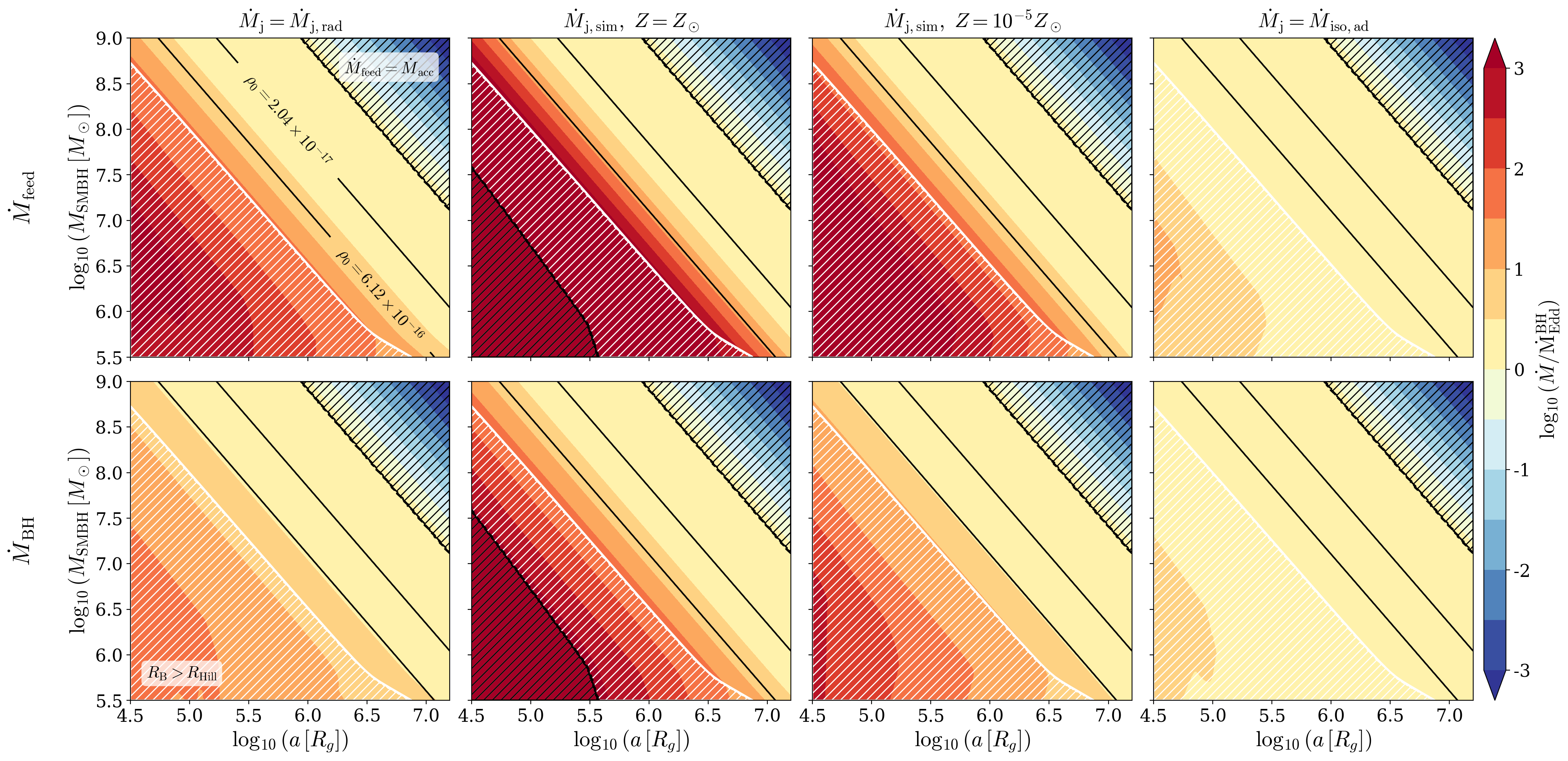}
 \caption{Cocoon regulated feeding rate $\dot{M}_{\rm feed}$ (top row) and the resulting black hole growth rate $\dot{M}_{\rm BH}$ (bottom row) in units of $\dot{\rm M}_{\rm Edd}^{\rm BH}$ across the ($M_{\rm SMBH}$, $a$) plane for \citetalias{2003sirko}-type disks with $\dot{M}_{\rm AGN} = \dot{\rm M}_{\rm Edd}^{\rm SMBH}$. Each column adopts one of the cocoon-regulated jet-flux prescriptions of Fig.~\ref{fig:mdotjet_5col_maps} -- from left to right, the radiative momentum-balance limit (Eq.~\ref{eq:mom_driven_regulated}), the fitted simulation scalings at solar and at low metallicity (\S\ref{subsec:results_sim_suite}), and the isotropic adiabatic limit (Eq.~\ref{eq:mdot_regulated}). White hatching marks the sheared interior, $R_{\rm B} > R_{\rm Hill}$, excluded from our analysis. Black hatching, bounded by the dashed line, marks where regulation fails -- the required feeding rate exceeds the available supply, $\dot{M}_{\rm feed} = \dot{M}_{\rm acc}$. Solid black contours trace the mid-plane densities sampled by our simulation suite, $\rho_0 = 2.0\times10^{-17}$ and $6.1\times10^{-16}\,{\rm g\,cm^{-3}}$.}
    \label{fig:mfeed_mbh_disk}
\end{figure*}

% ----------------------------------------------------------------------------------------------------
% sec:discussion
% subsec:disk_mdotjet
% ----------------------------------------------------------------------------------------------------

\subsection{Regulated jet mass flux across the disk}\label{subsec:disk_mdotjet}

Our simulations constrain the mass flux relaunched by the jet, $\dot M_j$, more directly than any other quantity. The regulated state is one in which the cocoon's outward momentum balances the inflow at the Bondi radius, and it is the outflow that adjusts to close that balance, while the rate reaching the BH follows only through the assumed mass loading (\S\ref{subsec:setup_accretion}). We therefore begin by mapping $\dot M_j$ alone across the outer disk, and turn to the implied feeding and growth rates in \S\ref{subsec:BH_growth} once that foundation is in place.

Fig.~\ref{fig:mdotjet_5col_maps} shows $\dot{M}_{\rm j}$, in units of $\dot{\rm M}_{\rm Edd}^{\rm BH}$, across the \citetalias{2003sirko} model for an AGN disk mass flux $\dot{M}_{\rm AGN} = \dot{\rm M}_{\rm Edd}^{\rm SMBH}$ and our fiducial $v_j = 10^4\, \rm \, km\, s^{-1}$. Because the marginally stable outer disk fixes the mid-plane density as the same function of $M_{\rm SMBH}$ and $a$ and opacity regulation holds the temperature near a common value, our results are insensitive to the choice of $\dot{M}_{\rm AGN}$. In each panel, $\dot{M}_{\rm j}$ is constructed by applying a different prescription for the suppression of $\dot{M}_{\rm j}$. Hatching marks the two boundaries of the region where our framework applies: interior to $R_{\rm B} = R_{\rm Hill}$ (white) the neglected tidal shear shapes the accretion flow, while beyond $\dot{M}_{\rm acc} = \dot{\rm M}_{\rm Edd}^{\rm BH}$ (gray) the outer disk accretion rate is already sub-Eddington and the trapping that powers the outflow shuts off.

The leftmost panel shows the expectation for $\dot{M}_j$ in the absence of any cocoon, i.e. the trapping prescription of \S\ref{subsec:super-Eddington_accretion}, in which the only limit on the outflow is the fraction of the inflow that the trapped-radiation region ejects. Because $f_{\rm eject}\rightarrow1$ wherever the supply is significantly super-Eddington (Fig.~\ref{fig:accretion_maps}), this prescription relaunches essentially the entire captured mass flux, $\dot{M}_{\rm j,trap}\simeq\dot{M}_{\rm acc}$. In Eddington units the predicted jet mass flux accordingly reaches $\sim10^{3}$--$10^{5}\,\dot{\rm M}_{\rm Edd}^{\rm BH}$ -- a maximally mass-loaded outflow carrying, in effect, the full Bondi--Hill supply.

The remaining panels replace this assumption with the cocoon-regulated mass flux scalings laid out in \S\ref{subsec:bondi_momentum}, \S\ref{subsec:momentum_driven} and \S\ref{subsec:results_sim_suite}. These are, in order, the momentum-driven limit [Eq.~\eqref{eq:mom_driven_regulated}], the fitted simulation scalings at sub-solar and solar metallicity, and the isotropic adiabatic limit [Eq.~\eqref{eq:mdot_regulated}, upper branch]. When written in units of the available mass supply $\dot{M}_{\rm acc}$, the two analytic prescriptions depend on local disk conditions only through $c_s$, reducing to a single number in the outer disk where temperatures are approximately constant,  $\dot M_j/\dot M_{\rm acc}\approx10^{-3.5}$ in the momentum-driven limit and $\approx10^{-6.3}$ in the adiabatic one at the fiducial jet velocity. The uniformity is a property of the ratio rather than of the flux itself, since $\dot M_{\rm acc}$ still climbs steeply with $\rho_0$ and written in the same units, the fitted scalings still carry the measured density dependence ($\propto \rho_0^{0.42}$ at solar metallicity, $\propto\rho_0^{0.12}$ at low). Because those measured dependences are steeper than the analytic ones, the ordering seen in Fig.~\ref{fig:suite_grid} does not persist across the full disk. There the momentum-regulated estimate lies above every simulated point; here, extrapolated inward in the AGN disk to higher densities and temperatures, the simulation-calibrated panels climb above it. The crossing is a direct consequence of the slope mismatch quantified in \S\ref{subsec:results_sim_suite}, and is the principal reason the two calibrated panels reach higher peak fluxes than either analytic limit.

The consequence, visible across the four rightmost panels, is a jet mass flux that is dramatically reduced relative to the trapping-only prediction. In the region of interest, exterior to $R_{\rm B}=R_{\rm Hill}$ but interior to $\dot{M}_{\rm acc}=\dot{\rm M}_{\rm Edd}^{\rm BH}$, $\dot{M}_j$ straddles the Eddington rate. Under the adiabatic constraint, it remains sub-Eddington, $\dot{M}_{\rm j}\sim10^{-2}$--$10^{-1}\, \dot{\rm M}_{\rm Edd}^{\rm BH}$, while in the radiative limit it climbs above Eddington as the mass supply grows toward the $R_{\rm B}=R_{\rm Hill}$ locus. The simulation-calibrated panels reach higher peak fluxes -- but the conclusion, that the jet mass flux is regulated to within roughly an order of magnitude of the Eddington rate across the Bondi-limited outer disk, is common to every cocoon-regulated prescription. 

% ----------------------------------------------------------------------------------------------------
% sec:discussion
% subsec:BH_growth
% ----------------------------------------------------------------------------------------------------

\subsection{Inferred BH Growth Rates}\label{subsec:BH_growth}

Our simulations launch the jet with a fixed mass-loading $\dot{M}_{\rm j} = \eta_{\rm m,fb}\,\dot{M}_{\rm BH}$ with $\eta_{\rm m,fb}\simeq1$, so that half of the captured flux is relaunched (an effective in-simulation ejected fraction of $f_{\rm eject} = 0.5$). The regulated state, however, is insensitive to this choice -- the cocoon admits whatever mass flux is required to hold the momentum balance at its boundary. In this section we therefore abandon the fixed loading and compute $\eta_{\rm m,fb}$ and $f_{\rm eject}$ self-consistently, taking the jet flux $\dot{M}_{\rm j}$ as the quantity our simulations constrain and identifying the jet with the outflow driven by the super-Eddington CBD. The relaunched flux is then $\dot{M}_{\rm j} = f_{\rm eject}\,\dot{M}_{\rm feed}$, with $f_{\rm eject}$ given by Eq.~\ref{eq:feject} except with the capture rate $\dot{M}_{\rm acc}$ replaced by the smaller, self-regulated feeding rate $\dot{M}_{\rm feed}$, which we now define. 

Before we proceed, it is useful to introduce a new variable $\dot{M}_{\rm feed}$ distinct from the capture rate $\dot{M}_{\rm acc}$. The latter is a property of the ambient disk alone -- fixed by the local density, sound speed and the smaller of the Bondi and Hill radii -- and is the flux that would reach the BH absent feedback. The outflow and resulting cocoon displace inflowing gas and admit only the fraction that momentum balance permits. We define this reduced flux delivered to the trapping region the feeding rate, $\dot M_{\rm feed}$. Given $\dot{M}_{\rm j}$, $\dot M_{\rm feed}$ then follows from inverting Eq.~\eqref{eq:feject} with $p=1/2$, evaluated against the feeding rate rather than~$\dot{M}_{\rm acc}$,
\begin{equation}
    \dot{M}_{\rm feed} - \left(\tfrac{5}{3}\,\dot{M}_{\rm feed}\, \dot{\rm M}_{\rm Edd}^{\rm BH}\right)^{1/2} = \dot{M}_{\rm j},
\label{eq:mfeed_inversion}
\end{equation}
a quadratic in $\dot{M}_{\rm feed}^{1/2}$, with the residual accreting onto the black hole,
\begin{equation}
    \dot{M}_{\rm BH} = \dot{M}_{\rm feed} - \dot{M}_{\rm j}
    = \left(\tfrac{5}{3}\,\dot{M}_{\rm feed}\,
    \dot{\rm M}_{\rm Edd}^{\rm BH}\right)^{1/2}.
\label{eq:mbh_residual}
\end{equation}
Solving the quadratic gives $\dot M_{\rm feed}$ in closed form,
\begin{equation}
    \dot M_{\rm feed} = \frac{1}{4}
    \left[\left(\tfrac{5}{3}\dot{\rm M}^{\rm BH}_{\rm Edd}\right)^{1/2}
    + \left(\tfrac{5}{3}\dot{\rm M}^{\rm BH}_{\rm Edd}
    + 4\dot M_j\right)^{1/2}\right]^{2}.
\label{eq:mfeed_solution}
\end{equation}
Two limits bound this construction. Where the required
$\dot{M}_{\rm feed}$ would exceed $\dot{M}_{\rm acc}$, the jet cannot regulate the inflow and we revert to $\dot{M}_{\rm feed} = \dot{M}_{\rm acc}$. Where $\dot{M}_{\rm feed} < \tfrac{5}{3}\dot{\rm M}_{\rm Edd}^{\rm BH}$ the trapping radius falls inside the inner disk edge and the full feeding rate accretes, $\dot{M}_{\rm BH} = \dot{M}_{\rm feed} = \dot{M}_{\rm acc}$. 
In this construction the ejected fraction and mass loading are no longer imposed but follow from the regulated state itself: $f_{\rm eject} = 1 - (\tfrac{5}{3}\,\dot{\rm M}^{\rm BH}_{\rm Edd}/\dot{M}_{\rm feed})^{1/2}$ rises from zero at the trapping threshold to $\simeq 0.8$--$0.97$ at the inner boundary $R_{\rm B}=R_{\rm Hill}$ (for the radiative-limit and solar-metallicity scalings, respectively), corresponding to effective mass loadings $\eta_{\rm m,fb} = \dot{M}_{\rm j}/\dot{M}_{\rm BH}$ between $0$ and $\sim 30$, in place of the fixed $\eta_{\rm m,fb} \simeq 1$ of the simulations. 

Fig.~\ref{fig:mfeed_mbh_disk} maps the resulting feeding (top row) and growth (bottom row) rates across the same disk model as Fig.~\ref{fig:mdotjet_5col_maps}, for the four cocoon-regulated prescriptions. The white hatching again marks the sheared interior, $R_{\rm B} > R_{\rm Hill}$, while the black hatching marks where regulation fails, $\dot{M}_{\rm feed} = \dot{M}_{\rm acc}$, so that the suppression reverts to the trapping-only expectation of \S\ref{subsec:super-Eddington_accretion}. This occurs not only in the outer, sub-Eddington corner common to all panels, but also -- for the solar metallicity simulation scaling alone -- at low $M_{\rm SMBH}$ and $a$, where the fitted jet flux, with its steep density dependence, demands more mass flux than the disk supplies. 

Between these boundaries, the feeding rate is close to the Eddington rate in every prescription. Its floor is set exactly by the trapping threshold, $\dot{M}_{\rm feed} \geq \tfrac{5}{3}\dot{\rm M}_{\rm Edd}^{\rm BH}$, at the outer edge of the regulated region, and its maximum is reached at the inner boundary $R_{\rm B}=R_{\rm Hill}$: approximately $1.6\times10^{3}\,\dot{\rm M}_{\rm Edd}^{\rm BH}$ for the solar-metallicity scaling -- the steepest density dependence and hence the largest excursion -- compared with $40\,\dot{\rm M}_{\rm Edd}^{\rm BH}$ for the radiative limit, while the isotropic adiabatic limit pins the feed to almost precisely the Eddington value throughout. The corresponding growth rates (bottom row), compressed by the square root in Eq.~\ref{eq:mbh_residual}, span only $\dot{M}_{\rm BH} \sim (1\text{--}50)\,\dot{\rm M}_{\rm Edd}^{\rm BH}$ across all four prescriptions -- an order of magnitude (or more) below the trapping-only expectation of \S\ref{subsec:super-Eddington_accretion}. The implication is that over most of the Bondi-limited outer disk, with the exception of a narrow region at the inner boundary of our models' validity, the accretion rate is suppressed to near-Eddington values, so that mechanical self-regulation largely resolves the overgrowth problem that motivates the Eddington caps of analytic disk models. It does so, moreover, continuously: in the cyclic models of \citet{2022tagawa}, where the jet breaks out of the disk and the evacuated cavity periodically refills, the duty cycle leaves the time-averaged rate well above Eddington, whereas the confined cocoon maintains the suppression throughout. Near the inner boundary $R_{\rm B}=R_{\rm Hill}$, where the growth rates approach $\sim 50\,\dot{\rm M}_{\rm Edd}^{\rm BH}$, and interior to it, the overgrowth problem remains open.

Finally, the dependence on jet velocity can be propagated analytically. Within the regulated region the feed is dominated by the jet term, $\dot{M}_{\rm feed}\simeq\dot{M}_{\rm j}$, so Eq.~\ref{eq:mbh_residual} gives $\dot{M}_{\rm BH}\propto\dot{M}_{\rm j}^{1/2}\propto v_{\rm j}^{-0.75}$ for the fitted $v_{\rm j}^{-1.5}$ suppression: a jet ten times slower raises the growth rate by a factor of $\sim6$, while leaving the qualitative picture -- growth pinned within an order of magnitude of Eddington -- unchanged. 

% ----------------------------------------------------------------------------------------------------
% sec:discussion
% subsec:disk_breakout
% ----------------------------------------------------------------------------------------------------

\subsection{Cocoon breakout versus confinement}\label{subsec:disk_breakout}

\begin{figure}
    \centering
\includegraphics[width=0.98\linewidth]{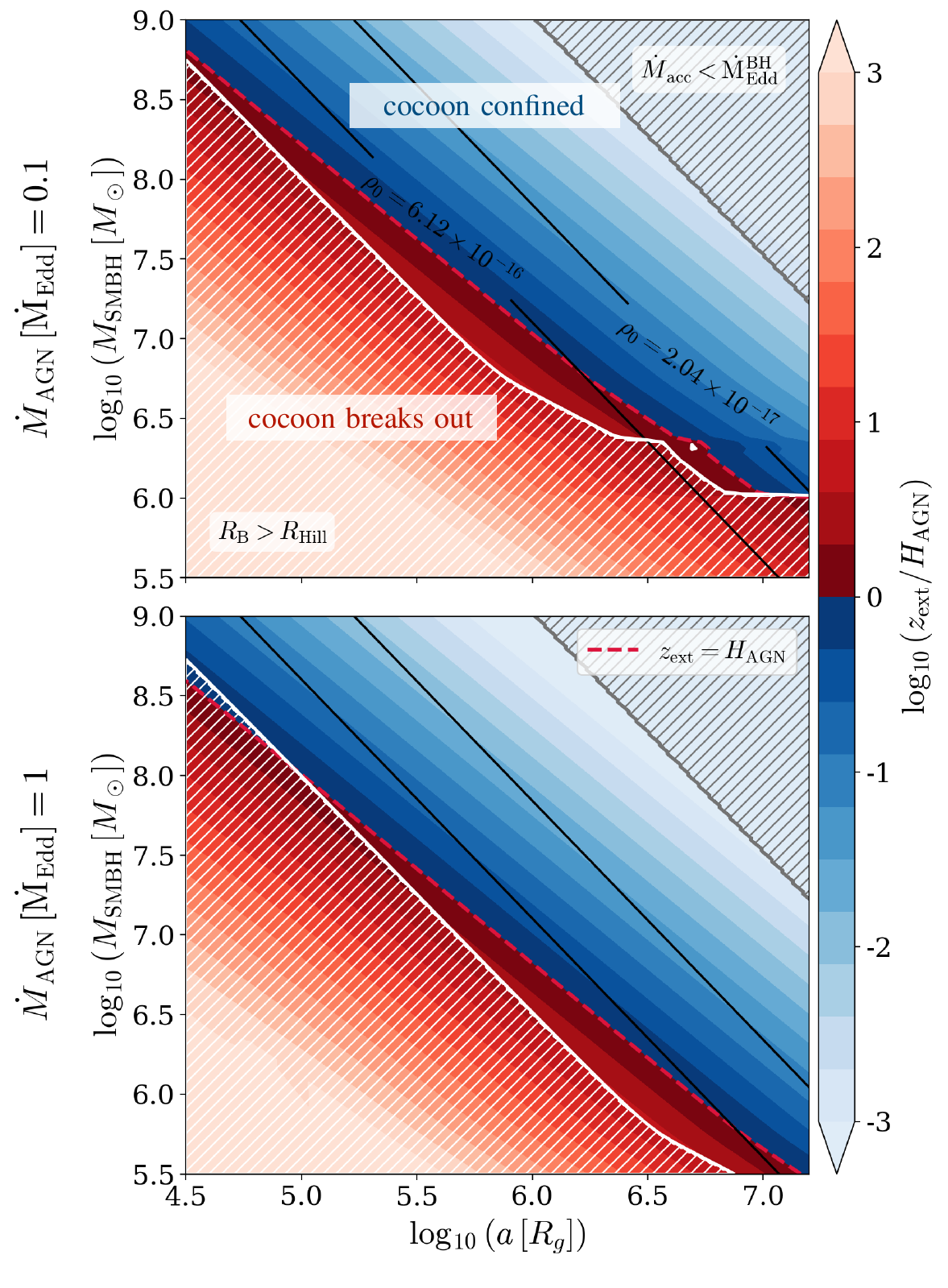}
 \caption{Estimated vertical extent of the jet cocoon relative to the local disk thickness, $z_{\rm ext}/H_{\rm AGN}$, across the \citetalias{2003sirko} disk plane of central mass $M_{\rm SMBH}$ and orbital distance $a$, for disk mass fluxes $\dot{M}_{\rm AGN} = 0.1$ (top) and $1\,\dot{\rm M}_{\rm Edd}^{\rm SMBH}$ (bottom). The vertical extent is computed as $z_{\rm ext} = (z/R)\,R_{\rm ext}$, with the aspect ratio and hot-phase radius taken from the fitted solar-metallicity scaling relations of \S\ref{subsec:results_sim_suite} evaluated at the fiducial jet velocity, $v_{\rm j}=10^{9}\,{\rm cm\,s^{-1}}$, and $H_{\rm AGN}$ from the disk model. The diverging color scale is darkest at the breakout boundary $z_{\rm ext} = H_{\rm AGN}$ (dashed crimson line), with confined cocoons ($z_{\rm ext} < H_{\rm AGN}$) in blue and vented cocoons in red. White hatching marks the sheared interior, $R_{\rm B} > R_{\rm Hill}$; gray hatching marks where the supply is sub-Eddington, $\dot{M}_{\rm acc} < \dot{\rm M}_{\rm Edd}^{\rm BH}$. Solid black contours trace the mid-plane densities sampled by our simulation suite (labeled in
the top panel).}
    \label{fig:zext_disk}
\end{figure}

Whether the cocoon remains buried in the disk or breaks out bears on both the regulated jet flux and the growth rates inferred from it. Once a cocoon crosses the disk surface the interior expands freely into the low density gas above the disk and depressurizes, removing the over-pressure at $R_{\rm B}$ on which the regulation depends. Regulation would then operate intermittently, through cycles of breakout and refilling, and the regulated rates mapped above would no longer apply \citep{2022tagawa}. Observationally, breakout would channel the jet's power out of the disk rather than depositing it locally, a potential source of transient emission from the AGN disk surface. Locating the breakout boundary is therefore a consistency check on the confined-feedback picture as well as a prediction in its own right. 

Figure~\ref{fig:zext_disk} maps the cocoon's vertical extent relative to the local disk thickness, $z_{\rm ext}/H_{\rm AGN}$, across the disk plane for our two disk models with $\dot{M}_{\rm AGN} = 0.1$\, or $1 \, \dot{\rm M}_{\rm Edd}^{\rm SMBH}$. Unlike the accretion-rate estimates of the preceding subsections, the breakout boundary cannot be anchored to an analytic prediction. Momentum regulation fixes the mass flux that the cocoon admits, but says nothing about the size the cocoon reaches; conversely, the expansion laws of \S\ref{sec:theory} that do predict a size assume free self-similar growth, which our simulations do not exhibit. There is therefore no analytic breakout criterion to compare against, and the estimate rests entirely on the fitted scalings of \S\ref{subsec:results_sim_suite}, $z_{\rm ext} = (z/R)\,R_{\rm ext}$, with both factors evaluated from the solar-metallicity fits at the fiducial jet velocity.

Because the fitted radius remains pinned near the Bondi radius and the aspect ratio is nearly constant, $z_{\rm ext}$ approximately tracks $R_{\rm B}$ throughout the AGN disk region of interest. The breakout condition $z_{\rm ext} = H_{\rm AGN}$ then reduces, to within the modest fitted elongation, to a comparison of the Bondi radius with the disk scale height. In general, wherever $R_{\rm B} < R_{\rm Hill}$, $R_{B}>H_{\rm AGN}$, so that the breakout boundary lies at or interior to the $R_{\rm B} = R_{\rm Hill}$ locus. If breakout occurs at all it occurs in the disk interior, in precisely the sheared, tidally dominated region our uniform-background simulations do not capture. 

The estimate is robust to the modelling choices at hand. The two disk models (top and bottom panels) yield nearly identical boundaries, differing appreciably only at low central masses, $M_{\rm SMBH}\lesssim10^{6}\,M_\odot.$ We show only the solar-metallicity scalings, since the low-metallicity fits are similar (\S\ref{subsec:results_sim_suite}) and metallicities of order solar are the relevant case for AGN disks. The jet velocity likewise has almost no effect. The fitted $v_j$ exponents of the radius and aspect ratio are individually small and largely cancel in the product, so maps at $v_j = 3\times10^{3}$ and $3\times10^{4}\,{\rm km\,s^{-1}}$ are nearly indistinguishable from Figure~\ref{fig:zext_disk}.

% ----------------------------------------------------------------------------------------------------
% sec:discussion
% subsec:caveats
% ----------------------------------------------------------------------------------------------------

\subsection{Caveats}

Several simplifications limit the direct application of these results. First, the jet mass-loading $\eta_{\rm m,fb}$ is held fixed within each run, whereas the trapping argument that sets its value implies it should fall as the cocoon drives the rate toward Eddington. Coupling $\eta_{\rm m,fb}$ to the regulated rate would introduce a stabilising back-reaction we do not capture, and runs with a dynamically varying loading will be needed to locate the resulting balance. Second, the uniform background omits the Keplerian shear of the disk, which is expected to distort the cocoon and reshape the accretion region wherever the Hill radius falls below the Bondi radius, so that a shearing-box treatment with a sub-Keplerian headwind will be required to test whether shear alters the momentum balance or only the cocoon morphology. Third, the outflow is launched as a single-velocity bipolar jet, whereas radiation-hydrodynamic models of super-Eddington outflows show a fast axial funnel giving way to a slower, mass-loaded wind at wider angles \citep{2021kitaki}, and an angle-dependent launch would test how sensitively the regulated rate depends on the assumed geometry. Fourth, radiative losses are followed through tabulated cooling rates without explicit radiation transport, so the opacity-dependent cooling of the shocked gas, the contribution of radiation pressure to driving the outflow, and the transition between optically thick and thin regions are not captured -- an omission that is most acute in the dense, optically thick inner disk, and that bears directly on the metallicity dependence we survey, since the opacity scales with metal content. Each of these simplifications is most consequential in the inner disk, where the accretion region approaches the Hill radius and the disk scale height and where radiation trapping is strongest. They therefore bear least on our conservative outer-disk estimates, and we leave their fuller treatment to future work.

% ----------------------------------------------------------------------------------------------------
% sec:conclusions
% ----------------------------------------------------------------------------------------------------

\section{Conclusions} \label{sec:conclusions}

We have carried out three-dimensional hydrodynamic simulations of the jet-driven cocoons that regulate accretion onto a stellar-mass black hole embedded in the dense gas of an AGN disk. Using \textsc{gizmo} with a particle-spawned jet whose power scales with the captured mass flux, we followed the gravitational gas capture, jet launching, and cocoon propagation around a $10\,\rm M_\odot$ black hole across a suite of runs spanning a range of jet velocities, ambient temperatures and densities under three cooling treatments -- an adiabatic run and radiative runs bracketing metallicities from $10^{-5}\,Z_\odot$ to solar.

Across this suite the self-regulated accretion rates are set by a momentum balance at the Bondi radius rather than by gravitational capture alone. In the adiabatic run the cocoon relaxes to a quasi-steady, mixing-regulated state, its hot interior holding a marginal overpressure $P_h\simeq2P_0$ that is fixed by accretion feedback, with the injected energy exported through a turbulent mixing layer in the absence of any radiative channel. The radiative runs neither retain their injected energy like an adiabatic bubble nor shed it like a fully radiative shell, but straddle the two regimes, radiating efficiently at the turbulent cocoon surface while holding a strongly overpressured interior, so that their time-averaged mass fluxes fall between the adiabatic and momentum-driven regulated rates.

At low metallicity, the cocoon develops a large-amplitude burst cycle of inflation, shell detachment, and refill that is a close mechanical analogue of the radiation-regulated accretion duty cycle of \citet{2009milosavljevic_a, 2009milosavljevic_b, 2011park, 2012park}. The solar-metallicity run shows only a weak, damped version of the same cycle, so that it is the cooling efficiency, set here by metallicity, that selects whether the strong, bursty mode or the mild, quasi-steady mode appears. In neither case does the cocoon expand self-similarly. Its radius remains pinned near the Bondi radius and its aspect ratio develops a modest, nearly constant elongation across the entire suite, departing from both the energy-conserving and momentum-driven expansion laws.

Within these limitations, our scalings allow a first estimate of self-regulated accretion across the outer disk. The simulations constrain the relaunched jet mass flux more directly than any other quantity. Absent a cocoon, the trapped-radiation region would relaunch essentially the full Bondi--Hill supply, giving $\dot M_j \sim (10^{3}$--$10^{5})\,\dot{\rm M}^{\rm BH}_{\rm Edd}$. Every cocoon-regulated prescription we tested instead holds it within roughly an order of magnitude of the Eddington rate across the Bondi-limited outer disk region, a suppression of three to five decades. 

Assuming wind losses inside the trapping radius predicted by radiatively inefficient super-Eddington disk models converts the bounds on $\dot{M}_j$ into a BH feeding rate and a growth rate $\dot{M}_{\rm BH}$. The feeding rate is pinned near Eddington throughout, with a floor set by the trapping threshold $\dot M_{\rm feed} \geq \tfrac{5}{3}\dot{\rm M}^{\rm BH}_{\rm Edd}$ and a maximum of $\sim 1.6\times10^{3}\,\dot{\rm M}^{\rm BH}_{\rm Edd}$ reached only at the inner boundary $R_{\rm B} = R_{\rm Hill}$, and the growth rates that follow are compressed by the square root of that relation into the narrow band $\dot M_{\rm BH} \sim (1\text{--}50)\,\dot{\rm M}^{\rm BH}_{\rm Edd}$. Over most of the Bondi-limited outer disk the accretion rate is therefore suppressed to near-Eddington values, largely resolving the overgrowth problem -- and doing so continuously, rather than through the breakout-and-refill duty cycle of \citet{2022tagawa}, which leaves the time-averaged rate well above Eddington. The exception is a narrow region at the inner boundary of our models' validity, $R_{\rm B} = R_{\rm Hill}$, where the maximum rates would still permit runaway growth toward intermediate masses; there, and interior to it, the problem remains open.

Across the same region we find that the cocoon does not expand self-similarly toward  breakout but stays confined near the Bondi radius, so that the finite vertical extent of the disk does not relieve the mass flux suppression. Because $R_{\rm B} < H_{\rm AGN}$ wherever $R_{\rm B} < R_{\rm Hill}$, the breakout boundary lies at or interior to the $R_{\rm B} = R_{\rm Hill}$ locus: if breakout occurs at all, it occurs in the sheared, tidally dominated inner disk that our uniform-background simulations do not capture, and assessing it will require the shearing-box treatment described above. Confinement carries an observational consequence. A cocoon that never crosses the disk surface deposits its energy locally rather than channeling it out of the disk, so we do not expect a solitary embedded BH to produce an optical or infrared transient of the kind proposed for merger-driven outflows; any escaping signature would have to be carried by radiation penetrating enough to traverse the overlying disk (hard X-rays or $\gamma$-rays). We stress that these conclusions inherit the simplifications listed above -- most importantly the fixed mass loading, the omission of Keplerian shear, and the absence of explicit radiation transport -- and that they are quantitatively anchored to fitted scalings whose underlying mechanism we do not yet fully explain. Taken together, however, they point in a single direction: feedback from an embedded black hole is confined, self-regulating, and sufficient to hold its growth near the Eddington rate throughout the outer disk, so that the embedded population should emerge from an AGN episode close to the mass with which it entered.

\section*{Acknowledgements}
We thank Hiromichi Tagawa, Brian Metzger, and Yuri Levin for thoughtful conversations and insights. KYS was supported by NASA through grants 80NSSC22K1124 and 80NSSC24K1224. RP acknowledges support from NASA award 80NSSC25K7554. ZH acknowledges financial support from NASA grant 80NSSC22K0822. The simulations are run on Frontera under allocation AST22010.

%%%%%%%%%%%%%%%%%%%%%%%%%%%%%%%%%%%%%%%%%%%%%%%%%%
\section*{Data Availability}
The data underlying this article will be shared on reasonable request to the corresponding author.



\bibliographystyle{mnras}
\bibliography{refs} % if your bibtex file is called example.bib

%\begin{thebibliography}{99}
%\end{thebibliography}

%%%%%%%%%%%%%%%%%%%%%%%%%%%%%%%%%%%%%%%%%%%%%%%%%%

%%%%%%%%%%%%%%%%% APPENDICES %%%%%%%%%%%%%%%%%%%%%
\appendix

\appendix
\section{Analytic expansion laws for energy-conserving bubbles and jet cocoons}
\label{app:bubble_scalings}

In this Appendix, we summarize a set of analytic similarity solutions used to interpret the evolution of pressure-driven bubbles (\S~\ref{app:weaver_solution}) and anisotropic cocoons (\S~\ref{app:anisotropic_uniform}). 

\subsection{Energy-conserving isotropic bubble}
\label{app:weaver_solution}

Consider a spherically symmetric, steady wind expanding at highly supersonic velocity into a uniform medium of density $\rho_0$. The wind drives a forward shock into the ambient gas and is itself thermalized at an inner reverse shock. Between the reverse shock and the contact discontinuity, the shocked wind forms a hot, low-density, approximately isobaric cavity. In the energy-conserving limit, radiative losses from this hot interior are negligible, and the evolution is governed by momentum conservation for the swept-up shell together with energy conservation for the hot bubble interior \citep{1977weaver}.

The shell is accelerated by the interior pressure acting over the spherical area $4\pi R^2$:
\begin{equation}
\frac{d}{dt}\left(M_{\rm sh}\dot R\right) = 4\pi R^2P,
\label{eq:app_iso_momentum}
\end{equation}
where $R(t)$ is the bubble radius and $P(t)$ is the interior pressure. For a uniform ambient medium, the swept-up shell mass is
\begin{equation}
M_{\rm sh} = \frac{4\pi}{3}\rho_0R^3.
\label{eq:app_iso_shell_mass}
\end{equation}
Neglecting cooling losses, the thermal energy of the interior obeys
\begin{equation}
\frac{dE_{\rm th}}{dt} = \dot E_{\rm in} - P\frac{dV}{dt},
\label{eq:app_iso_energy}
\end{equation}
where $\dot E_{\rm in} = \epsilon \dot{M}_j v_j^2$/2 is the mechanical luminosity injected by the wind and $V = \frac{4\pi}{3}R^3$. The interior thermal energy is related to pressure by
\begin{equation}
E_{\rm th} = \frac{PV}{\gamma-1} = \frac{4\pi}{3}\frac{R^3P}{\gamma-1}.
\label{eq:app_iso_eth}
\end{equation}

We seek a self-similar solution of the form
\begin{equation}
R(t)=R_0t^\alpha,
\qquad
P(t)=P_0t^\beta .
\end{equation}
Substituting into Eq.~\eqref{eq:app_iso_momentum} yields
\begin{equation}
P_0 = \frac{\rho_0}{3}\alpha(4\alpha-1)R_0^2,
\label{eq:app_iso_p0_mom}
\end{equation}
and matching powers of time gives
\begin{equation}
\beta=2\alpha-2.
\label{eq:app_iso_beta_mom}
\end{equation}
Energy conservation requires $PV\propto t$, so
\begin{equation}
\beta+3\alpha=1.
\label{eq:app_iso_beta_energy}
\end{equation}
Combining Eqs.~\eqref{eq:app_iso_beta_mom} and \eqref{eq:app_iso_beta_energy} gives
\begin{equation}
\alpha=\frac{3}{5},
\qquad
\beta=-\frac{4}{5}.
\end{equation}

The full normalization follows by inserting the solution into the energy equation. For $\gamma=5/3$, one obtains
\begin{align}
R(t)
&=
\left(\frac{125}{154\pi}\right)^{1/5}
\left(\frac{\dot E_{\rm in}}{\rho_0}\right)^{1/5}
t^{3/5},
\label{eq:app_weaver_R}
\\
P(t)
&=
\frac{5}{22\pi}
\left(\frac{125}{154\pi}\right)^{-3/5}
\dot E_{\rm in}^{2/5}\rho_0^{3/5}t^{-4/5}.
\label{eq:app_weaver_P}
\end{align}

\subsection{Energy-conserving anisotropic cocoon in a homogeneous medium}
\label{app:anisotropic_uniform}

We now consider an anisotropic, jet-driven cocoon. We model the cocoon as a cylinder of radius $R(t)$ and height $H(t) = 2 z(t)$, with volume
\begin{equation}
V= 2\pi R^2z.
\end{equation}
The jet injects mechanical power $\dot E_{\rm in}$ and thrust $\dot\Pi_j = \dot{M_j}v_j$. The axial head advances by balancing the jet thrust against the ram pressure of the ambient medium over an effective head area $A_h\simeq \pi R^2$:
\begin{equation}
\rho_0 A_h\dot z^2
\simeq
\frac{\dot\Pi_j}{2}.
\label{eq:app_aniso_head}
\end{equation}
Equivalently,
\begin{equation}
\dot z = \frac{K}{R},
\qquad
K\equiv
\left(\frac{\dot\Pi_j}{2\pi\rho_0}\right)^{1/2}.
\label{eq:app_K_uniform}
\end{equation}
The lateral expansion is driven by the hot-cocoon pressure acting on the cylindrical sidewall. We write the swept-up shell mass as
\begin{equation}
M_{\rm sh} = 2 \pi\,\rho_0 R^2z,
\label{eq:app_aniso_mass}
\end{equation}
The lateral momentum equation is then
\begin{equation}
\frac{d}{dt}\left(M_{\rm sh}\dot R\right)= 
4\pi RzP .
\label{eq:app_aniso_lat_mom}
\end{equation}
The interior thermal energy obeys
\begin{equation}
    \frac{d}{dt}\left(\frac{PV}{\gamma-1}\right) = \dot E_{\rm in} + P\dot V_\perp,
\label{eq:app_aniso_energy}
\end{equation}
where the pressure-driven lateral volume growth is
\begin{equation}
    \dot V_\perp = 4\pi Rz\dot R.
\end{equation}
This expression separates the pressure-driven lateral expansion from the axial advance of the jet head. The latter increases the cocoon volume, and hence dilutes the interior energy, but the axial work is supplied directly by the jet thrust rather than by the isotropic cocoon pressure.

As in \S~\ref{app:weaver_solution}, we derive a self-similar solution
\begin{equation}
    R(t)=A t^{2/3},
    \qquad
    z(t)=B t^{1/3},
    \qquad
    P(t)=D t^{-2/3}.
\label{eq:app_aniso_ansatz}
\end{equation}
The Eq.~\eqref{eq:app_K_uniform} gives
\begin{equation}
    B=\frac{3K}{A},
\label{eq:app_aniso_B}
\end{equation}
and the lateral momentum equation gives
\begin{equation}
    D =\frac{4}{9}\rho_0A^2.
\label{eq:app_aniso_D}
\end{equation}
Finally, the energy equation gives
\begin{equation}
    \dot E_{\rm in} = \frac{17}{6}\pi A^2BD.
\label{eq:app_aniso_energy_norm}
\end{equation}
Combining Eqs.~\eqref{eq:app_aniso_B}--\eqref{eq:app_aniso_energy_norm}, we find
\begin{equation}
    A =
    \left[
    \frac{9}{68}
    \frac{\dot E_{\rm in}}{\pi\rho_0K}
    \right]^{1/3}.
    \label{eq:app_aniso_A_general}
\end{equation}
Substituting $\dot{E}_{\rm in} = \epsilon \dot{M}_j v_j^2/2$ and $\dot{\Pi}_{\rm j} = \dot{M}_j v_j$  yields the scalings 
\begin{align}
    R(t)
    &\propto
    \epsilon^{2/3}
    \dot M_j^{1/6}
    v_j^{1/2}
    \rho_0^{-1/6}
    t^{2/3},
    \label{eq:app_aniso_R}
    \\
    z(t)
    &\propto
    \epsilon^{-1/3}
    \dot M_j^{1/3}
    \rho_0^{-1/3}
    t^{1/3},
    \label{eq:app_aniso_z}
    \\
    P(t)
    &\propto
    \epsilon^{2/3}
    \dot M_j^{1/3}
    v_j
    \rho_0^{2/3}
    t^{-2/3}.
    \label{eq:app_aniso_P}
\end{align}
%

%If the cocoon pressure collimates the jet, the effective head area is instead set by the recollimation shock. Following the scaling arguments of \citet{1989begelman} and \citet{2011bromberg}, one may write
%
%\begin{equation}
%A_h
%\sim
%\chi_B\theta_0^2
%\frac{\dot\Pi_j}{P_c},
%\label{eq:app_collimated_head_area}
%\end{equation}
%
%where $\theta_0$ is the injected half-opening angle and $\chi_B$ is an order-unity geometric factor. Substitution into Eq.~\eqref{eq:app_K_uniform} gives $\dot z\sim \theta_0^{-1}(P_c/\rho_0)^{1/2}$, while lateral expansion proceeds at $\dot R\sim(P_c/\rho_0)^{1/2}$. Thus a pressure-confined jet evolves with an approximately fixed aspect ratio, $z/R\sim\theta_0^{-1}$, and obeys the energy-conserving scalings
%
%\begin{equation}
%R\propto t^{3/5},
%\qquad
%z\propto t^{3/5},
%\qquad
%P_c\propto t^{-4/5}.
%\label{eq:theory_collimated_scalings}
%\end{equation}
%

\section{Testing simulation dependence on sub-grid accretion prescription and resolution} 
\label{app:t_disk_res}

\begin{figure}
    \centering
\includegraphics[width=0.30\textwidth]{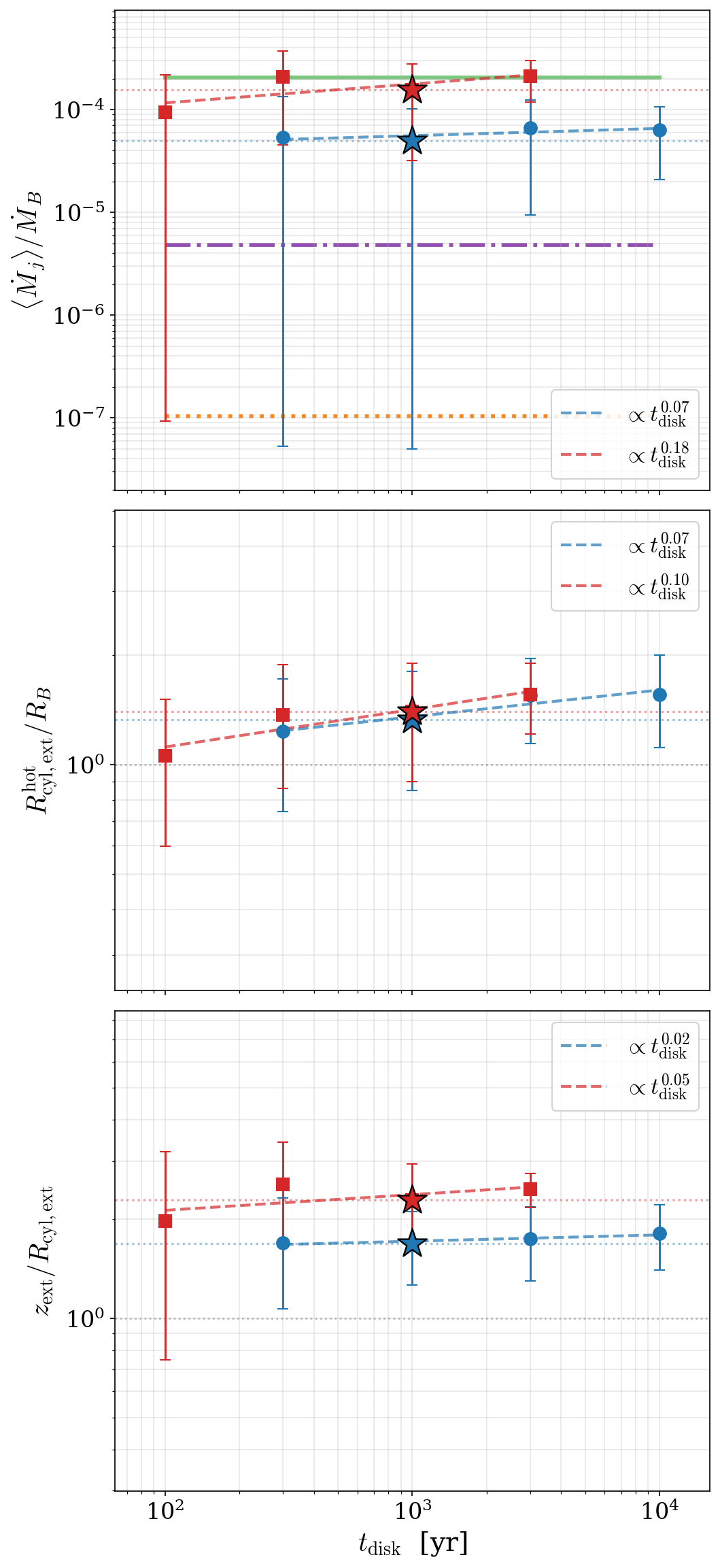}
 \caption{Dependence of the regulated state on the sub-grid draining time $t_{\rm disk}$, for low-metallicity ($Z=10^{-5}\,Z_\odot$, blue circles) and solar-metallicity ($Z=Z_\odot$, red squares) runs. Rows show the time-averaged accretion rate $\langle\dot{M}_{\rm j}\rangle/\dot{M}_{\rm B}$ (top), the radial extent of the hot phase $R^{\rm hot}_{\rm cyl,ext}/R_{\rm B}$ (middle), and the aspect ratio $z/R^{\rm hot}_{\rm cyl,ext}$ (bottom). Points are time averages over each run and error bars span the minimum and maximum values, with power-law fits to each metallicity subset shown dashed and labelled in each panel. In the top row we also mark the analytic reference scalings, the momentum-regulated expectation (green), the isotropic-adiabatic estimate $\dot{M}_{\rm iso,ad}$ (purple), and the cylindrical-adiabatic estimate $\dot{M}_{\rm cyl,ad}$ (orange). All three quantities are nearly independent of $t_{\rm disk}$.}
    \label{fig:tacc_test}
\end{figure}

\begin{figure}
    \centering
\includegraphics[width=0.3\textwidth]{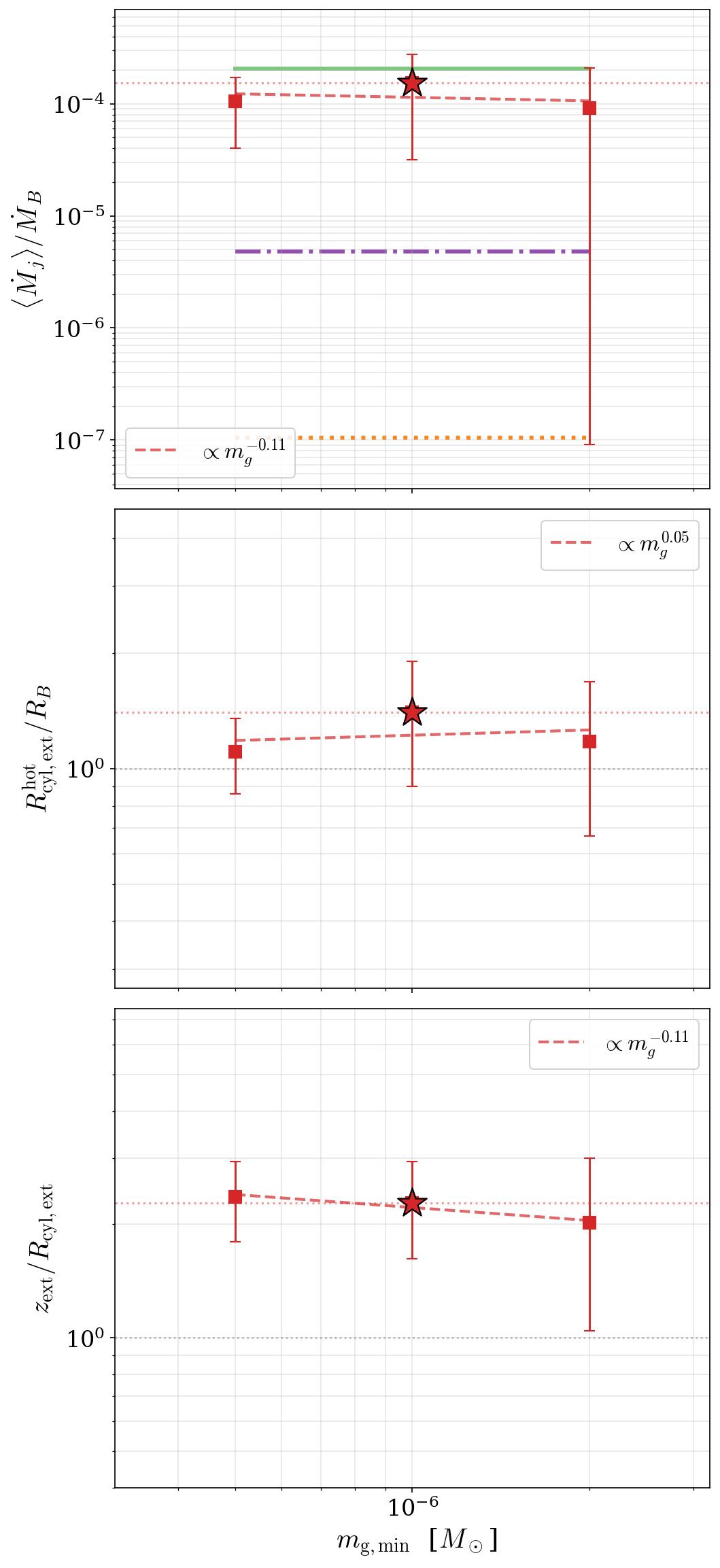}
 \caption{As Fig.\ref{fig:tacc_test}, for the minimum gas mass resolution $m_{\rm g,min}$ at solar metallicity ($Z=Z_\odot$, red squares). Rows show the time-averaged accretion rate $\langle\dot{M}_{\rm j}\rangle/\dot{M}_{\rm B}$ (top), the radial extent of the hot phase $R^{\rm hot}_{\rm cyl,ext}/R_{\rm B}$ (middle), and the aspect ratio $z/R^{\rm hot}_{\rm cyl,ext}$ (bottom), with power-law fits (dashed) and analytic reference scalings as in Fig.\ref{fig:tacc_test}. The regulated state is unchanged across a factor of four in particle mass.}
    \label{fig:res_test}
\end{figure}

Our fiducial suite fixes the sub-grid draining time at $t_{\rm disk}=10^3$~yr and the minimum gas mass at $m_{\rm g,min}=10^{-6}\,M_\odot$. We confirm here that neither choice alters the regulated state. Fig.\ref{fig:tacc_test} varies $t_{\rm disk}$ over $10^2$--$10^4$~yr at both metallicities, and Fig.\ref{fig:res_test} varies $m_{\rm g,min}$ over $5\times10^{-7}$--$2\times10^{-6}\,M_\odot$ at solar metallicity. In both sweeps the time-averaged accretion rate, the cocoon radius, and the aspect ratio are nearly flat, with fitted slopes well below those of the jet velocity and effective temperature in \S\ref{subsec:results_sim_suite}, and every point lies within the error bars of the rest. The shortest draining times raise the variability of $\dot{M}_{\rm acc}$, as the smaller reservoir drains in fewer, larger increments, but the time-averaged rate and the cocoon structure are unchanged. We therefore hold $t_{\rm disk}$ and $m_{\rm g,min}$ fixed across the main suite, consistent with the weak draining-time dependence reported by \citet{2023su}.

%%%%%%%%%%%%%%%%%%%%%%%%%%%%%%%%%%%%%%%%%%%%%%%%%%

% Don't change these lines
\bsp	% typesetting comment
\label{lastpage}
\end{document}